\documentclass[aps,twocolumn,groupedaddress]{revtex4}
\usepackage{epsfig}
\usepackage{graphicx}
\usepackage[T1]{fontenc}
\usepackage{ae}
\usepackage{color}
\usepackage[latin1]{inputenc}
\usepackage{amssymb,amsbsy,amsmath}
\usepackage{bbm}

\newtheorem{lemma}{Lemma}
\newtheorem{defi}{Definition}
\newtheorem{prop}{Proposition}
\newtheorem{ass}{Assumption}
\newtheorem{theorem}{Theorem}
\newtheorem{cor}{Corollary}

\begin{document}



\title[Theory of ground states I]
{Theory of ground states for classical Heisenberg spin
systems I}

\author{Heinz-J\"urgen Schmidt$^1$
\footnote[1]{Correspondence should be addressed to hschmidt@uos.de}
}
\address{$^1$Universit\"at Osnabr\"uck, Fachbereich Physik,
Barbarastr. 7, D - 49069 Osnabr\"uck, Germany}


\begin{abstract}
We formulate part I of a rigorous theory of ground states for classical, finite, Heisenberg spin systems.
The main result is that all ground states can be constructed from the eigenvectors of a real, symmetric matrix
with entries comprising the coupling constants of the spin system as well as certain Lagrange parameters. The eigenvectors
correspond to the unique maximum of the minimal eigenvalue considered as a function of the Lagrange parameters.
However, there are rare cases where all ground states obtained in this way have unphysical dimensions $M>3$ and the theory
would have to be extended. Further results concern the degree of additional degeneracy, additional to the trivial degeneracy
of ground states due to rotations or reflections. The theory is illustrated by a couple of elementary examples.
\end{abstract}

\maketitle

\section{Introduction}\label{sec:I}
In quantum mechanics the ground states of a system are the
eigenvectors of the Hamiltonian H corresponding to the lowest energy
eigenvalue. Thus there is a clear recipe how to find ground states:
Just diagonalize the Hamiltonian. In practice this may turn out to
be numerically difficult, nevertheless it is a straightforward
procedure. The analogous problem for a classical Heisenberg spin
system cannot be solved in an analogous fashion.
Although the definition of ground states is clear (states where the
classical Hamiltonian assumes its minimum) the ground states can
only be analytically determined in special cases.
Numerical procedures are available, but they
may converge slowly, and provide no guarantee that the obtained state represents a global, not only local
minimum of energy. After all, one is never sure that
the numerical procedures will find all ground states, which may be crucial for calculations of thermodynamic
properties at low temperatures.
It may be that there are additional ground states that cannot be obtained from the known ones
by rotations or reflections.
Another problem is the dimensionality of ground states. Under which conditions there exist
1-dimensional, 2- dimensional or only 3-dimensional ground states?
The latter problem is also connected with ``frustration":
Classical spin systems are frustrated if they do not possess 1-dimensional ground states
(but not vice versa, see example 5 in section \ref{sec:EU}).
Existing theories mainly focus
on spin lattices, see the seminal work of Luttinger and Tisza \cite{LT46}, followed by \cite{LK60}, \cite{L74}, \cite{FF74}
and the more recent publications \cite{N04},  \cite{XW13} based on this approach. An alternative
approach is \cite{SL03}, but this is mainly focussed on finite systems with large
point group symmetries and does not cover the general case.

Hence there is the need for a general theory of classical ground
states that settles the mentioned problems.  I will try to outline such a
theory although a couple of questions will remain open. Since this theory exceeds the format of
a single article I have decided to split it into different parts of which the present
paper will be the first one.

The first four sections after this Introduction contain general results illustrated by elementary examples whereas the
proofs are given in a separate section \ref{sec:P}. This makes it, hopefully, possible to obtain a general
survey without dwelling upon mathematical details. The mathematics used in the proofs is rather elementary
and presumably known to all physicists with a moderate background in mathematics. One exception
might be the use of certain concepts of convex analysis that are not much common in a physical context (except
in the foundations of quantum mechanics). Here we have to refer the reader to the pertinent literature, e.~g., to \cite{R97}.
We will summarize the central results of this paper in a theorem \ref{TSUM}, see section \ref{sec:SUM},
that contains also the pertinent definitions and can be read independently of the main text.

The method we adopt to tackle the ground state problem I have dubbed the ``Lagrange variety approach",
see subsection \ref{sec:DLV}. It is based on the observation that the ground states satisfy the ``stationary state equation" (SSE)
involving certain Lagrange parameters due to the constraints of constant spin lengths.
The SSE can be cast into the form of an eigenvalue equation for some matrix that we call the
``dressed ${\mathbbm J}$-matrix". Its entries are the coupling constants between the spins in the Heisenberg model plus certain
Lagrange parameters ${\boldsymbol\lambda}$ in the diagonal. The set of eigenvalues of the dressed ${\mathbbm J}$-matrix depending
on ${\boldsymbol\lambda}$ is called the ``Lagrange variety" ${\mathcal V}$. In this way we obtain a $1:1$ connection between the solutions of the
SSE and certain points of ${\mathcal V}$, called ``elliptic points". In section \ref{sec:SP} we give a geometrical characterization
of the elliptic points of ${\mathcal V}$ that essentially says that in an infinitesimal neighborhood of these points
the Lagrange variety is given by the surface of a ``vertical double cone", see Figure \ref{FIGEX2} for an illustration.
For the minimal eigenvalue of the dressed ${\mathbbm J}$-matrix
there exists a unique point of ${\mathcal V}$ with a vertical double cone and hence a ground state living on the corresponding
eigenspace, see section \ref{sec:EU}.
In the special case where the minimal eigenvalue has a smooth maximum we obtain a $1$-dimensional ground state.
However, it may happen that all ground states obtained in this way will be
$M$-dimensional, $M>3$, and hence unphysical. In this case one has to look for other elliptic points of ${\mathcal V}$ in order
to find physical ground states. We provide an example in section \ref{sec:NS}.
Nevertheless, these examples are rare in practice and the approach of the present paper seems to be useful.

This approach also gives interesting results for the problem of degeneracy, see subsection \ref{sec:DAD}.
All ground states of Heisenberg systems are trivially degenerate in the sense that arbitrary rotations/reflections are always possible.
But sometimes ``additional degeneracy" occurs, for example, if independent rotations of a subset of spin vectors are possible.
The theory tells us how the degree of additional degeneracy can be read off from any ground state of maximal dimension.
One simple example is the anti-ferromagnetic bow tie that can be viewed as resulting from the ``fusion" of two triangles
and shows an additional degeneracy of degree one, see subsection \ref{sec:DAD}.
The general process of fusion is sketched in subsection \ref{sec:DF}. If we also admit unphysical ground states with $M>3$
it can be shown that no further degeneracy occurs, i.~e.~, all ground states have the same Lagrange parameters, see subsection \ref{sec:EU}.

This has important practical consequences.
Assume that we consider a certain Heisenberg spin system and look for ground states.
As mentioned above there exist simple codes to numerically determine certain ground states. For example, we can
start with a random $3$-dimensional spin configuration and fix a certain spin number $\mu=1,\ldots,N$.
Then we choose the spin vector ${\mathbf s}_\mu$ such that the energy of the interaction of the spin $\mu$ with all other spins is minimized.
We consider the next spin $\mu+1$ and repeat the process until the change of the total energy is smaller than a given $\epsilon>0$.
If the repetition of the whole  procedure with different inial conditions
gives reproducible results we can be rather sure that we have found some ground states.
But how to find all ground states?
Application of the present theory suggests to calculate the Lagrange parameters of the numerically determined ground state and to
examine the eigenvalue and the corresponding eigenspace of the dressed ${\mathbbm J}$-matrix. If the eigenvalue is minimal
(and this will be the typical case), we have no problems with unphysical ground states:
We can easily solve the ``additional degeneracy equation" (ADE), see subsection \ref{sec:DAD}, and
thus find all additional degeneracies, provided the degree of additional degeneracy is not too large.
Some of these ground states may be unphysical, but all physical ones are included.
Then we are done: The theory tells us that there are no further ground states.

After having outlined the content of the present paper with the optimistic number I in its title
it will be in order to say a few words about possible extensions that may be covered by forthcoming papers.
Besides the Lagrange variety approach there exists another approach that I will call ``Gram set approach".
Its main idea is to linearize the energy function that is bilinear in the spin vectors,  analogous
to the linearization of the expectation value in quantum mechanics by the introduction of ``statistical operators".
The operator analogous to the statistical operator is the ``Gram matrix" defined in subsection \ref{sec:DAD}.  The Gram set approach
is not a substitute for the Lagrange variety approach but a supplement that deepens the understanding of the ground state problem.
Further, it will be useful to illustrate the complete solution of the ground state problem for the general classical spin triangle.

In the present paper we have mainly provided elementary examples where the set of ground states was already known in order to
illustrate our theory, example 5 in section \ref{sec:EU} being an exception.
What is still missing are more applications to systems where the complete set of ground states
is either completely unknown or only partially known. A possible candidate for the latter case is the anti-ferromagnetic cuboctahedron,
where additional degeneracy due to independent rotations has been found \cite{S10}.

Another question is to what extent the present results could be generalized to
spin systems where the Hamiltonian is no longer of Heisenberg type, but, say, still bilinear in the spin components.
This would include dipole-dipole interactions as well as corrections of Dzyaloshinsky-Moriya type.
Whereas the first steps following the SSE can be accordingly generalized, see, e.~g., \cite{SSL16}, I am pessimistic about the
possibility to generalize central parts of the theory to non-Heisenberg systems.

But there is a special case of non-Heisenberg Hamiltonians that is particularly interesting for physical applications,
namely a Heisenberg Hamiltonian plus a Zeeman term describing the interaction of the spins with an
outer magnetic field $B$. This case in some sense can be traced back to the pure Heisenberg case.
First, one observes that in the presence of a magnetic field the ground states will be among the ``relative ground states",
i.~e.~, ground states for a given total spin $S$. The latter satisfy an analogous SSE with an additional Lagrange parameter, say, $\alpha$
due to the additional constraint $S^2=$ const.~. The terms involving $\alpha$ can be distributed to the dressed ${\mathbbm J}$-matrix
in such a way that one obtains an SSE of the pure Heisenberg form and the present theory can be applied. The only difference is
that the entries of the dressed ${\mathbbm J}$-matrix proportional to $\alpha$ have a different physical meaning
and $\alpha$ is not a given constant but may vary over some domain.
At any case, the extension of the present theory to the case of $B\neq 0$ seems to be highly desirable.

Another realm of possible future work would be the specialization of the present theory to cases with a large symmetry group
and the comparison to the known results of \cite{LT46}--\cite{XW13} or \cite{SL03}.
A few remarks about the symmetric case already can be found in section \ref{sec:EU} as well as a Theorem \ref{TheoremSym}
about the existence of symmetric ground states.
Since we have assumed finite spin systems from the outset
an application to infinite spin lattices could probably only be made in the sense of approximating the lattice by a finite system
with periodic boundary conditions.

\section{General definitions and results}\label{sec:D}
\subsection{The Lagrange variety approach}\label{sec:DLV}

The classical phase space $\mathcal P$ for the systems of $N$ spins under consideration
consists of all configurations of spin vectors (or ``states")
\begin{equation}\label{D1}
\mathbf{s}_\mu,\;\mu=1,\ldots,N
\;,
\end{equation}
subject to the constraints
\begin{equation}\label{D2}
\mathbf{s}_\mu\cdot \mathbf{s}_\mu=1,\;\mu=1,\ldots,N
\;.
\end{equation}
From a physical point of view one is only interested in those cases where the vectors occurring in (\ref{D1}) and (\ref{D2})
are at most $3$-dimensional. However, this restriction turns out to be mathematically unnatural and hence will be cancelled.
Thus the vectors occurring in (\ref{D1}) and (\ref{D2}) are assumed to be elements of $\mathbb{R}^M$ where $M$ is some natural number
that may assume different values throughout the paper. The corresponding phase space ${\mathcal P}_M$ is the $N$-fold product
of unit spheres
\begin{equation}\label{DPh}
 {\mathcal P}_M\equiv S^{M-1}\times\ldots\times S^{M-1}
\end{equation}
and hence compact. We will use the natural embeddings ${\mathcal P}_M \subset {\mathcal P}_M'$ for $M<M'$.
Extending the dimension of spin vectors for mathematical reasons does not mean that
we ignore the fact that in physical applications this dimension must not exceed $3$.
We have still the possibility to retrieve the physical spin configurations from a larger set of mathematical configurations by looking at their dimensions.
The exact definition of ``dimension" is given in the following paragraph.

Let ${\mathbf s}$ denote the $N\times M$-matrix with entries ${\mathbf s}_{\mu,i},\;\mu=1,\ldots,N,\;i=1,\ldots,M$.
According to the different use of Greek and Latin indices it will be always clear that ${\mathbf s}_\mu$ denotes the $\mu$-th row of
${\mathbf s}$ and ${\mathbf s}_i$ its $i$-th column.
The ``dimension" $\text{dim}({\mathbf s})$ of ${\mathbf s}$ is simply defined as its matrix rank. Hence it is equal to the maximal
number of linearly independent rows  ${\mathbf s}_\mu$ of ${\mathbf s}$, or, equivalently, to the maximal number of linearly independent
columns ${\mathbf s}_i$ of ${\mathbf s}$. It follows that always $\mbox{dim } ({\mathbf s})\le N$.
 According to the physical parlance we will speak of ``collinear states"
or "Ising states" in case of $\text{dim}({\mathbf s})=1$, and ``co-planar states" in case of $\text{dim}({\mathbf s})=2$. The case of
$\text{dim}({\mathbf s})=3$ has not yet received a particular denomination and will be referred to as ${\mathbf s}$ being a ``$3$-dimensional state".

The Heisenberg Hamiltonian $H$ is a smooth function defined on ${\mathcal P}_M$ of the form
\begin{equation}\label{D3}
H(\mathbf{s})=\sum_{\mu,\nu=1}^N J_{\mu \nu}\,\mathbf{s}_\mu\cdot \mathbf{s}_\nu
\;,
\end{equation}
where the coupling coefficients $J_{\mu \nu}$ are considered as the entries of a real, symmetric $N\times N$ matrix $\mathbb J$
with vanishing diagonal.

The Hamiltonian (\ref{D3}) does not uniquely determine the symmetric matrix $\mathbb J$: Let $\lambda_\mu,\,\mu=1,\ldots,N$
be arbitrary real numbers subject to the constraint
\begin{equation}\label{D4}
\sum_{\mu=1}^N \lambda_\mu =0
\;,
\end{equation}
and define a new matrix ${\mathbbm J}({\boldsymbol\lambda})$ with entries
\begin{equation}\label{D5}
J({\boldsymbol\lambda})_{\mu\nu}\equiv J_{\mu\nu}+\delta_{\mu\nu}\lambda_\mu
\;,
\end{equation}
then
\begin{eqnarray}\label{D6a}
\tilde{H}({\mathbf s})&\equiv& \sum_{\mu,\nu=1}^N J({\boldsymbol\lambda})_{\mu \nu}\,\mathbf{s}_\mu\cdot \mathbf{s}_\nu\\
\label{D6b}
&=&
\sum_{\mu,\nu=1}^N J_{\mu \nu}\,\mathbf{s}_\mu\cdot \mathbf{s}_\nu+
\sum_{\mu=1}^N \lambda_\mu\,\mathbf{s}_\mu\cdot \mathbf{s}_\mu\\
\label{D6c}
&=& H({\mathbf s})
\;,
\end{eqnarray}
due to (\ref{D2}) and (\ref{D4}). The transformation
$J_{\mu\nu} \rightarrow J({\boldsymbol\lambda})_{\mu\nu}$ according to (\ref{D5}) has been called a ``gauge transformation"
in \cite{SL03} according to the close analogy with other branches of physics where this notion is common.
In most problems the simplest gauge would be the ``zero gauge", i.~e.~, setting $\lambda_\mu=0$ for $\mu=1,\ldots,N$.
However, in the present context it is crucial not to remove the gauge freedom by a certain choice of the $\lambda_\mu$
but to retain it. We will hence explicitly stress the dependence of the coupling matrix on the undetermined
$\lambda_\mu$ by using the notation ${\mathbbm J}({\boldsymbol\lambda})$. ${\mathbbm J}({\boldsymbol\lambda})$ will be
called the ``dressed ${\mathbbm J}$-matrix" and its entries will be, as above, denoted by $J({\boldsymbol\lambda})_{\mu\nu}$.
 The rationale is that we want to trace back
the properties of ground states to the eigenvalues and eigenvectors of
${\mathbbm J}({\boldsymbol\lambda})$ and these in a non-trivial way depend on ${\boldsymbol\lambda}$.
The ``undressed" matrix ${\mathbbm J}$ without ${\boldsymbol\lambda}$ will always denote a symmetric $N\times N$-matrix in the zero gauge.\\
Let $\Lambda$ denote the $N-1$-dimensional subspace of ${\mathbb R}^N$  defined by
\begin{equation}\label{DLambda}
\Lambda\equiv \left\{{\boldsymbol\lambda}\in{\mathbb R}^N\left| \sum_{\mu=1}^N\,\lambda_\mu=0\right.\right\}
\end{equation}
As coordinates in $\Lambda$ we will use the first $N-1$ components $\lambda_i,\,i=1,\ldots,N$ since
the $N$-th component can be expressed by the others via $\lambda_N=-\sum_{i=1}^{N-1} \lambda_i$.\\

A ``ground state" of the spin system is defined as any configuration ${\mathbf s}\in{\mathcal P}_N$ where $H({\mathbf s})$ assumes its global minimum
$E_{min}$. We will also say that $\mathbf s$ is the ground state of the Hamiltonian $H$ or of ${\mathbbm J}$.
The restriction to ${\mathcal P}_N$ does not exclude any ground state of whatever dimension since always $\mbox{dim }({\mathbf s})\le N$.
The existence of ground states is guaranteed since the continuous function $H$ defined on the compact set ${\mathcal P}_N$
assumes its minimum at some points ${\mathbf s}$ of ${\mathcal P}_N$.
Let us define the set of ground states by
\begin{equation}\label{DGS1}
\breve{\mathcal P}  \equiv \left\{{\mathbf s}\in {\mathcal P}_N\left|H({\mathbf s})=E_{min}\right.\right\}.
\end{equation}
In general there exist a lot of ground states. For example, a global rotation or reflection of a ground state is again a ground state
due to the invariance of the Hamiltonian (\ref{D3}) under rotations/reflections.
The group of rotations/reflections $R$ of ${\mathbbm R}^M$ defined by the property $R^\top=R^{-1}$
is usually denoted by $O(M)$; hence we will also speak of $O(M)$-equivalence of ground states.
Later we will present examples that show additional degeneracies
of the ground states apart from the ``trivial" rotational/reflectional degeneracy.
If there is no additional degeneracy, i.~e.~, if any two ground states are $O(M)$-equivalent we
will also say that the ground state is ``essentially unique".
Let $\breve{M}$ be the maximal dimension of ground states, i.~e.~,
\begin{equation}\label{DGS2}
  \breve{M}\equiv \mbox{Max }\left\{\mbox{dim }({\mathbf s})\left|{\mathbf s}\in\breve{\mathcal P} \right. \right\}.
\end{equation}
It can be shown that for any ground state ${\mathbf s}\in\breve{\mathcal P}$ there exists an $R\in O(N)$ such that
$R\,{\mathbf s}\in {\mathcal P}_{\breve{M}}\subset {\mathcal P}_{N}$ w.~r.~t.~the above-mentioned natural embedding of phase spaces.
Hence we can always assume that ground states ${\mathbf s}$ are $N\times \breve{M}$-matrices. Nevertheless, it will be often more
convenient not to fix $M=\breve{M}$ but to use an undetermined integer $M$ in the pertinent definitions.\\

It is well-known that a smooth function of $M\times N$ variables has a vanishing gradient at those points where it assumes its
(local or global) minimum.
If the definition domain of the function is constrained, as in our case,
its gradient no longer vanishes at the minima but will only be perpendicular
to the ``constraint manifold". For a rigorous account see, e.~g., \cite{AMR83}.
The resulting  equation reads, in our case,
\begin{equation}\label{D7}
\sum_{\nu=1}^N J_{\mu\nu}{\mathbf s}_\nu = - \kappa_\mu\,{\mathbf s}_\mu,\quad \mu=1,\ldots,N
\;.
\end{equation}
Here the $\kappa_\mu$  are the Lagrange parameters due to the constraints (\ref{D2}). This equation is only
necessary but not sufficient for ${\mathbf s}$ being a ground state. If it is satisfied we call the corresponding state a "stationary state" and will
refer to (\ref{D7}) as the ``stationary state equation" (SSE). This wording of course reflects the fact that exactly the stationary states
will not move according to the equation of motion for classical spin systems, see, e.~g., \cite{SL03}, but we will not dwell upon this here.
All ground states are stationary states but there are stationary states that are not ground states.
Let us rewrite (\ref{D7}) in the following way:
\begin{equation}\label{D8}
\sum_{\nu=1}^N J_{\mu\nu}{\mathbf s}_\nu = (\bar{\kappa}- \kappa_\mu)\,{\mathbf s}_\mu-\bar{\kappa}\,{\mathbf s}_\mu
=-\lambda_\mu\,{\mathbf s}_\mu-\bar{\kappa}\,{\mathbf s}_\mu
\;,
\end{equation}
where we have introduced the mean value of the Lagrange parameters
\begin{equation}\label{D9}
\bar{\kappa}\equiv \frac{1}{N}\sum_{\mu=1}^N\,\kappa_\mu
\;,
\end{equation}
and the deviations from the mean value
\begin{equation}\label{D10}
\lambda_\mu\equiv \kappa_\mu-\bar{\kappa},\;\mu=1,\ldots,N
\;.
\end{equation}
We denote by $\Lambda_0\subset\Lambda$ the set of vectors
${\boldsymbol\lambda}$ with components (\ref{D10}) resulting from
(\ref{D7}) in the case of a ground state ${\mathbf s}\in\breve{\mathcal P}$. Later we will
prove that $\Lambda_0$ consists of a single point $\Lambda_0=\{\hat{\boldsymbol\lambda}\}$
but at the moment we will not use this fact.
${\boldsymbol\lambda}\in\Lambda_0$ will be called a ``ground state gauge".
It can be used for a gauge transformation $J_{\mu\nu}
\rightarrow J({\boldsymbol\lambda})_{\mu\nu}$ which renders (\ref{D8}) in the
form of an eigenvalue equation:
\begin{equation}\label{D11}
\sum_{\nu=1}^N J({\boldsymbol\lambda})_{\mu\nu}{\mathbf s}_{\nu} = -\bar{\kappa}\,{\mathbf s}_{\mu}
\;,
\end{equation}
or, in matrix form,
\begin{equation}\label{D12}
{\mathbbm J}({\boldsymbol\lambda}){\mathbf s} = -\bar{\kappa}\,{\mathbf s}
\;.
\end{equation}
This means that each column ${\mathbf s}_i,\; i=1,\ldots,M$ of the matrix $\mathbf s$ will be an eigenvector of the matrix
${\mathbbm J}({\boldsymbol\lambda}),\,{\boldsymbol\lambda}\in\Lambda_0$ corresponding to the eigenvalue $-\bar{\kappa}$.

Since this situation will occur very often throughout the paper we will use the abbreviating phrase
``$\varphi$ is an eigenvector of $(A,a)$"  iff
the eigenvalue equation $A\,\varphi = a\,\varphi$ holds for $\varphi\neq 0$.
We note that a global rotation/reflection
${\mathbf s}\mapsto {\mathbf s}'$ where
${\mathbf s}'_{\mu i} = \sum_{j=1}^M R_{ij}{\mathbf s}_{\mu j},\; R\in O(M),$
does not affect the eigenvalue $-\bar{\kappa}$ and the ground state gauge ${\boldsymbol\lambda}\in\Lambda_0$. In this sense, the
rotational/reflectional degeneracy is factored out by the present approach.

The connection between the minimal energy $E_{min}$  and the eigenvalue $-\bar{\kappa}$ is given by
\begin{eqnarray}\label{D13a}
E_{min}&=&\sum_{\mu,\nu=1}^N J_{\mu\nu}\,{\mathbf s}_\nu\cdot{\mathbf s}_\mu
\stackrel{(\ref{D6c})}{=}\sum_{\mu,\nu=1}^N J({\boldsymbol\lambda})_{\mu\nu}\,{\mathbf s}_\nu\cdot{\mathbf s}_\mu\\
\label{D13b}
 &\stackrel {(\ref{D11})}{=}& -\bar{\kappa} \sum_{\mu=1}^N
{\mathbf s}_\mu\cdot{\mathbf s}_\mu \stackrel {(\ref{D2})}{=}-N\,\bar{\kappa}
\;.
\end{eqnarray}

It will be instructive to consider the reverse process. Let ${\mathbf s}_i,\; i=1,\ldots n,$ be the eigenvectors
of  ${\mathbbm J}({\boldsymbol\lambda})$ for some ${\boldsymbol\lambda}\in\Lambda$ corresponding to an $n-$fold degenerate eigenvalue. Then the eigenvectors
need not lead to a spin configuration since $\sum_{i=1}^n {\mathbf s}_{\mu,i}^2$ may depend on $\mu$. If $\sum_{i=1}^n {\mathbf s}_{\mu,i}^2$
is independent of $\mu$ and hence can be taken as $1$, the spin vector will generally be $n-$dimensional. Even if $n\le 3$, we have only
obtained a stationary state that need not be a ground state. This illustrates the problems inherent to a general theory of ground states.

We introduce some more notation.
Let ${\boldsymbol\lambda}\in\Lambda$ be arbitrary and ${\mathbf s}\in\breve{\mathcal P}$ be any ground state of the spin system.
Further, let $j_\alpha({\boldsymbol\lambda})$ denote the $\alpha-$th eigenvalue of ${\mathbbm J}({\boldsymbol\lambda})$ and
$j_{min}({\boldsymbol\lambda})$ its lowest eigenvalue.
Application of the Rayleigh-Ritz variational principle to the present situation yields
\begin{eqnarray}\label{D14a}
E_{min}&=&\sum_{\mu,\nu=1}^N \sum_{i=1}^M J_{\mu\nu}\,{\mathbf s}_{\nu,i}\,{\mathbf s}_{\mu,i}\\
\label{D14b}
&\stackrel{(\ref{D6c})}{=}&\sum_{\mu,\nu=1}^N \sum_{i=1}^M J({\boldsymbol\lambda})_{\mu\nu}\,{\mathbf s}_{\nu,i}\,{\mathbf s}_{\mu,i}\\
\label{D14c}
&\ge& j_{min}({\boldsymbol\lambda}) \sum_{\nu=1}^N \sum_{i=1}^M {\mathbf s}_{\nu,i}^2 \stackrel{(\ref{D2})}{=} N j_{min}({\boldsymbol\lambda})
\;.
\end{eqnarray}
We stress that (\ref{D14a})-(\ref{D14c}) holds for every gauge ${\boldsymbol\lambda}\in\Lambda$, not only for a ground state gauge.
It seems plausible that for the ground state gauge, i,~e.~, for ${\boldsymbol\lambda}\in\Lambda_0$ the inequality (\ref{D14c})
can be replaced by an equality. This is indeed the case, see Theorem \ref{Theorem2},
and means that a ground state can be built from the eigenvectors of
$({\mathbbm J}({\boldsymbol\lambda}),j_{min}({\boldsymbol\lambda})),\,{\boldsymbol\lambda}\in\Lambda_0$.
But it may happen that all ground states obtained in this way have a dimension greater than $3$. We will present an example in section \ref{sec:NS}.
If this is not the case, that is, if $\breve{M}\le 3$
we define the spin system to be a ``standard" one.

From (\ref{D14a})-(\ref{D14c}) it follows that
$\frac{1}{N}E_{min}$ is an upper bound of the function $j_{min}:\Lambda\longrightarrow{\mathbb R}$.
We will show below that the function $j_{min}$ assumes its upper bound at some ${\boldsymbol\lambda}\in\Lambda$.

Let $p({\boldsymbol\lambda},x)=\det\left( {\mathbbm J}({\boldsymbol\lambda})-x\,{\mathbbm 1}\right)$ denote the characteristic polynomial of ${\mathbbm J}({\boldsymbol\lambda})$. The set
\begin{equation}\label{DdefV}
{\mathcal V}= {\mathcal V}({\mathbbm J})\equiv\left\{ ({\boldsymbol\lambda},x)\in \Lambda\times {\mathbb R}\,|\,p({\boldsymbol\lambda},x)=0 \right\}
\end{equation}
is a ``real algebraic variety", see, e.~g., \cite{CL07}
and will be called the ``Lagrange variety" of the classical spin system under consideration since the parameters
$({\boldsymbol\lambda},x)$ are in $1:1$ relation to the Lagrange parameters $\kappa_\mu,\,\mu=1,\ldots,N$
of the SSE (\ref{D7}). The graph of the function $j_{min}:\Lambda\longrightarrow{\mathbb R}$ is a subset of the
Lagrange variety. The points $({\boldsymbol\lambda},x)$ of ${\mathcal V}({\mathbbm J})$ can be divided into two disjoint subsets:
$({\boldsymbol\lambda},x)$ will be called ``singular" if the gradient $\nabla p({\boldsymbol\lambda},x)$  vanishes:
$\frac{\partial p({\boldsymbol\lambda},x)}{\partial x}=\frac{\partial p({\boldsymbol\lambda},x)}{\partial \lambda_i}=0$ for
$i=1,\ldots,N-1$. Otherwise, $({\boldsymbol\lambda},x)$ will be called ``regular". In the neighbourhood of a regular point
${\mathcal V}({\mathbbm J})$ will be a smooth $N-1$ dimensional manifold embedded into ${\mathbb R}^N$
and its tangent space at $({\boldsymbol\lambda},x)$ will be orthogonal
to $\nabla p({\boldsymbol\lambda},x)$. Note that the vanishing of $\frac{\partial p({\boldsymbol\lambda},x)}{\partial x}$ means that
the eigenvalue $x$ of ${\mathbbm J}({\boldsymbol\lambda})$ is at least doubly degenerate. In this case we are necessarily at a singular point of
${\mathcal V}({\mathbbm J})$:

\begin{prop}\label{PropV}
If $p({\boldsymbol\lambda},x)=\frac{\partial p({\boldsymbol\lambda},x)}{\partial x}=0$, then
$\frac{\partial p({\boldsymbol\lambda},x)}{\partial \lambda_i}=0$ for $i=1,\ldots,N-1$ and hence $({\boldsymbol\lambda},x)$
is a singular point of ${\mathcal V}({\mathbbm J})$.
\end{prop}

The proofs of this and following propositions and theorems will be given in a separate section \ref{sec:P}.

\begin{prop}\label{Prop1}
$j_{min}:\Lambda\longrightarrow{\mathbb R}$ is a concave function, i.~e.~,
$j_{min}(\alpha {\boldsymbol\lambda}+(1-\alpha){\boldsymbol\mu})\ge \alpha\, j_{min}({\boldsymbol\lambda})+(1-\alpha)j_{min}({\boldsymbol\mu})$
for all ${\boldsymbol\lambda},{\boldsymbol\mu} \in\Lambda$ and $\alpha\in [0,1]$.
\end{prop}

From this one concludes the following, see \cite{R97}, Cor.~10.1.1:
\begin{cor}\label{Cor1}
$j_{min}$ is a continuous function.
\end{cor}

Since the set $\{j_{min}({\boldsymbol\lambda})|{\boldsymbol\lambda}\in\Lambda\}$ is bounded from above by $\frac{1}{N}E_{min}$, see
(\ref{D14a})--(\ref{D14c}), its supremum $\hat{\jmath}\equiv\sup\,\{j_{min}({\boldsymbol\lambda})|{\boldsymbol\lambda}\in\Lambda\}$ exists.
It can be shown that $j_{min}$ assumes this supremum at some set $\widehat{J}$:
\begin{prop}\label{Prop2}
The set $\widehat{J}\equiv \{{\boldsymbol\lambda}\in\Lambda|j_{min}({\boldsymbol\lambda})=\hat{\jmath}\}$ is a non-empty
compact, convex subset of $\Lambda$.
\end{prop}

We close this subsection with an elementary example.

\vspace{5mm}
\fbox{Example 1: The dimer ($N=2$)}
\vspace{5mm}

\begin{figure}[ht]
  \centering
    \includegraphics[width=1.0\linewidth]{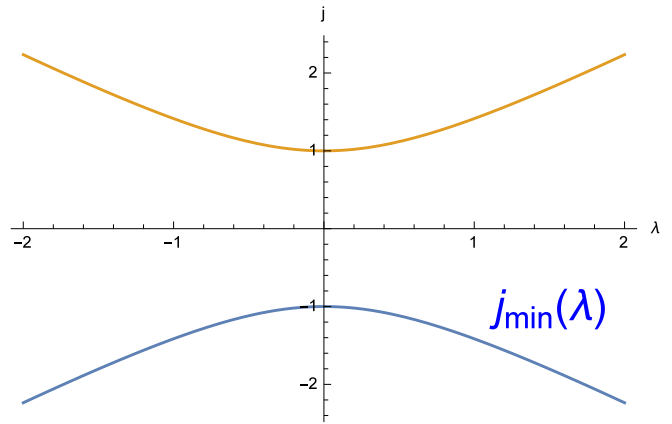}
  \caption[Example 1]
  {The Lagrange variety ${\mathcal V}$ of the AF dimer consists of two disjoint curves. The lower one
  is the graph of the function $j_{min}(\lambda)$.
  It has a smooth maximum at $\lambda=0$ corresponding to
  a collinear ground state $\uparrow\,\downarrow$.}
  \label{FIGEX1}
\end{figure}

In the antiferromagnetic (AF) case the matrices ${\mathbbm J}$ and ${\mathbbm J}(\lambda)$ assume the form
\begin{equation}\label{ex11}
 {\mathbbm J} = \left(\begin{array}{cc} 0 &  1\\ 1 & 0 \end{array}\right),\quad
 {\mathbbm J}(\lambda) = \left(\begin{array}{cc} \lambda &  1\\ 1 & -\lambda \end{array}\right)
 \;,
\end{equation}
and the characteristic equation of the latter is $\det\left(  {\mathbbm J}(\lambda)-x\,{\mathbbm 1}\right) =x^2-(1+\lambda^2)=0$.
It has the two solutions $x_\pm = \pm\sqrt{1+\lambda^2}$ and hence $j_{min}(\lambda)=-\sqrt{1+\lambda^2}$, see Figure \ref{FIGEX1}.

The function $j_{min}(\lambda)$ has a unique maximum at $\lambda=0$ of height $\hat{\jmath}=j_{min}(0)=-1$.
At this maximum the dressed ${\mathbbm J}$-matrix assumes the form
\begin{equation}\label{ex12}
 {\mathbbm J}(0) = \left(\begin{array}{cc}0 &  1\\ 1 & 0 \end{array}\right)
 \;,
\end{equation}
and has the eigenvector $\varphi=\left(\begin{array}{r} 1\\-1\end{array}\right)$ corresponding to the eigenvalue
$\hat{\jmath}=j_{min}(0)=-1$. This yields the collinear ground state
${\mathbf s}_1=1,\; {\mathbf s}_2=-1$, symbolically ${\mathbf s}=\uparrow\,\downarrow$.

In the ferromagnetic case $j_{min}(\lambda)$ is unchanged, but at its maximum the dressed ${\mathbbm J}$-matrix
assumes the form
\begin{equation}\label{ex13}
 {\mathbbm J}(0) = \left(\begin{array}{rr}0 &  -1\\ -1 & 0 \end{array}\right)
 \;,
\end{equation}
and has the eigenvector $\varphi=\left(\begin{array}{c} 1\\1\end{array}\right)$ corresponding to the eigenvalue
$\hat{\jmath}=j_{min}(0)=-1$. This yields the collinear ground state
${\mathbf s}_1=1,\; {\mathbf s}_2=1$, symbolically ${\mathbf s}=\uparrow\,\uparrow$.

\subsection{Degeneracy}\label{sec:DAD}

We will recapitulate and generalize some notions already introduced in \cite{SL03}.
As in the previous subsection let ${\mathbf s}$ denote the $N\times M$-matrix with entries ${\mathbf s}_{\mu,i},\;\mu=1,\ldots,N,\;i=1,\ldots,M$.
Let ${\mathbf s}^\top$ denote the transposed matrix.
For each ${\mathbf s}\in{\mathcal P}_M$ we define the ``Gram matrix" $G\equiv {\mathbf s}\,{\mathbf s}^\top$
with entries $G_{\mu\nu}={\mathbf s}_\mu\cdot{\mathbf s}_\nu$, where $\cdot$ denotes the usual inner product of ${\mathbb R}^M$.
Hence $G$ will be a symmetric $N\times N$-matrix that is positively semi-definite, $G\ge 0$, and satisfies $G_{\mu\mu}=1$
for all $\mu=1,\ldots,N$. Moreover, $\mbox{rank}(G)=\mbox{rank}({\mathbf s})\le M$.

Conversely, if $G$ is a positively semi-definite $N\times N$-matrix with rank $M\le N$, satisfying $G_{\mu\mu}=1$
for all $\mu=1,\ldots,N$. Then the spectral representation of $G$  yields
\begin{equation}\label{G1}
G= \sum_{i=1}^M \,\gamma_i \,{\mathbbm P}_{\varphi_i}
\;,
\end{equation}
where the $\gamma_i >0$ are the non-zero eigenvalues and  ${\mathbbm P}_{\varphi_i}$ denote the
projectors onto the corresponding unit eigenvectors ${\varphi_i}$ of $G$, $i=1,\ldots,M$.
Their matrix entries are given by
\begin{equation}\label{DAD1a}
\left( {\mathbbm P}_{\varphi_i}\right)_{\mu\nu}=\varphi_{i\mu}\,\varphi_{i\nu}\mbox{ for }\mu,\nu=1\ldots,N\;.
\end{equation}

Then we define $N$ spin vectors ${\mathbf s}_\mu\in{\mathbb R}^M$ with components
${\mathbf s}_{\mu i}=\sqrt{\gamma_i}\,\varphi_{i \mu}$ and conclude
\begin{equation}\label{G2}
{\mathbf s}_\mu\cdot{\mathbf s}_\nu = \sum_{i=1}^M {\mathbf s}_{\mu i}{\mathbf s}_{\nu i}
=\sum_{i=1}^M \gamma_i \varphi_{i \mu}\varphi_{i \nu}\stackrel{(\ref{DAD1a})(\ref{G1})}{=} G_{\mu\nu}
\;.
\end{equation}
Moreover, the ${\mathbf s}_\mu$ are unit vectors since ${\mathbf s}_\mu\cdot{\mathbf s}_\mu=G_{\mu\mu}=1$ for $\mu=1,\ldots,N$.

The correspondence between spin configurations ${\mathbf s}\in{\mathcal P}_M$ and Gram matrices $G$ is many-to-one: Let $R\in O(M)$,
then the two configurations ${\mathbf s}_\mu$ and $R\,{\mathbf s}_\mu,\;\mu=1,\ldots,N$ will obviously yield the same Gram matrix.
Actually, this is the only possibility where two configurations have the same $G$ according to the following

\begin{prop}\label{Prop5}
Let ${\mathbf s}^{(i)}\in{\mathcal P}_M,\;i=1,2,$
be two  spin configurations satisfying\\ ${\mathbf s}^{(1)}_\mu\cdot{\mathbf s}^{(1)}_\nu ={\mathbf s}^{(2)}_\mu\cdot{\mathbf s}^{(2)}_\nu$
for all $\mu,\nu=1,\ldots N$, then there exists a rotation/reflection $R\in O(M)$ such that ${\mathbf s}^{(2)}_\mu=R\,{\mathbf s}^{(1)}_\mu$
for all $\mu=1,\ldots N$.
\end{prop}
Hence the representation of spin configurations by Gram matrices exactly removes the ``trivial" rotational/reflectional degeneracy of possible ground states;
the set of Gram matrices is in $1:1$ correspondence with the set of $O(M)$-equivalence classes of states. We note in passing
that the energy $H({\mathbf s})$ of a spin configuration ${\mathbf s}$ may be written in a linearized form by using the Gram matrix as
$H({\mathbf s})=\mbox{Tr }\left( G\,{\mathbbm J}\right)$.

Next we want to give a more precise definition of the phrase that a spin configuration ${\mathbf s}$ can be built from the
vectors of some eigenspace $S$ of ${\mathbbm J}({\boldsymbol\lambda})$ or, equivalently, that ${\mathbf s}$ is ``living on $S$".
To this end we consider a general linear subspace $S\subset{\mathbb R}^N$ and define:
\begin{defi}\label{DAD1}

\begin{enumerate}
  \item $S$ is called ``$M$-elliptic" iff there exists an  $\mathbf{s}\in{\mathcal P}_M$ such that
  its columns ${\mathbf s}_i,\;i=1,\ldots,M$ are elements of $S$.
  \item If $S$ is $M$-elliptic we define \\
  ${\mathcal P}_{M,S}\equiv \{\mathbf{s}\in{\mathcal P}_M\left|{\mathbf s}_i\in S\mbox{ for all }i=1,\ldots,M\right.\}$.
  \item $S$ is called ``elliptic" iff it is $M$-elliptic for some integer $M\ge 1$.
  \item $S$ is called ``completely elliptic"  iff there exists an $\mathbf{s}\in{\mathcal P}_M$ such that
  its columns ${\mathbf s}_i,\;i=1,\ldots,M$ are elements of $S$, and moreover, $\mbox{dim } {\mathbf s}=\mbox{ dim } S=M$.
\end{enumerate}
\end{defi}

\vspace{5mm}
\fbox{Example 2}\\

\begin{figure}[ht]
  \centering
    \includegraphics[width=1.0\linewidth]{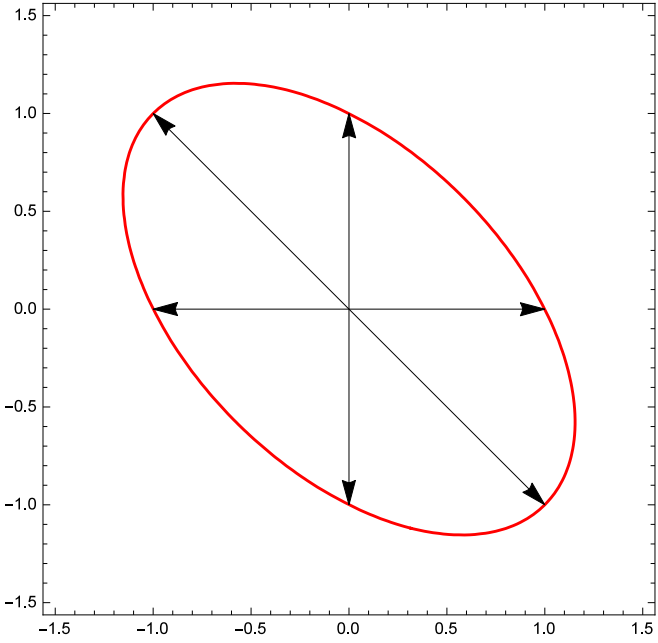}
  \caption[Example 1]
  {The six row vectors of the matrix  (\ref{ell2}) the two column vectors of which span an elliptic subspace.
  All six vectors lie on the (red) ellipse defined by $x^2+y^2+x\,y=1$.}
  \label{FIGELL}
\end{figure}

In order to illustrate the wording of Definition \ref{DAD1} we consider a system  of $N=6$ spins with ${\mathbbm J}$-matrix
\begin{equation}\label{ell1}
{\mathbbm J}=\left(
\begin{array}{rrrrrr}
 0 & 1 & 2 & -1 & -1 & 1 \\
 1 & 0 & -1 & 2 & -1 & 1 \\
 2 & -1 & 0 & 1 & 1 & -1 \\
 -1 & 2 & 1 & 0 & 1 & -1 \\
 -1 & -1 & 1 & 1 & 0 & 2 \\
 1 & 1 & -1 & -1 & 2 & 0 \\
\end{array}
\right)
\end{equation}
Its lowest eigenvalue is $j_{min}=-4$ with a two-dimensional eigenspace $S$ spanned by the column vectors of the matrix
\begin{equation}\label{ell2}
W= \left(
\begin{array}{rr}
 1 & 0 \\
 -1 & 1 \\
 -1 & 0 \\
 1 & -1 \\
 0 & 1 \\
 0 & -1 \\
\end{array}
\right).
\end{equation}
The six row vectors of $W$ lie on the ellipse $x^2+y^2+x\,y=1$, see Figure \ref{FIGELL}.
It can be shown that $S$ is also elliptic in the sense of the Definition \ref{DAD1}:
Defining
\begin{equation}\label{ell4}
{\boldsymbol\Gamma}\equiv\left(
\begin{array}{cc}
 \frac{\sqrt{2+\sqrt{3}}}{2} & \frac{\sqrt{2-\sqrt{3}}}{2} \\
 \frac{\sqrt{2-\sqrt{3}}}{2} & \frac{\sqrt{2+\sqrt{3}}}{2} \\
\end{array}
\right),
\end{equation}
we can show that another basis of $S$ is given by the column vectors of ${\mathbf s}=W\,{\boldsymbol\Gamma}$:
\begin{equation}\label{ell3}
{\mathbf s}=\left(
\begin{array}{cc}
 \frac{\sqrt{2+\sqrt{3}}}{2} & \frac{\sqrt{2-\sqrt{3}}}{2} \\
 -\frac{1}{\sqrt{2}} & \frac{1}{\sqrt{2}} \\
 -\frac{1}{2} \sqrt{2+\sqrt{3}} & \frac{1}{4}
   \left(\sqrt{2}-\sqrt{6}\right) \\
 \frac{1}{\sqrt{2}} & -\frac{1}{\sqrt{2}} \\
 \frac{\sqrt{2-\sqrt{3}}}{2} & \frac{\sqrt{2+\sqrt{3}}}{2} \\
 \frac{1}{4} \left(\sqrt{2}-\sqrt{6}\right) & -\frac{1}{2}
   \sqrt{2+\sqrt{3}} \\
\end{array}
\right),
\end{equation}
such that the six rows of ${\mathbf s}$ are unit vectors. \\

For general $M$-dimensional elliptic subspaces spanned
by the columns of some matrix $W$ the corresponding row vectors will lie on a central $M$-dimensional ellipsoid, in general not unique,
that can be transformed into a unit sphere by some linear symmetric transformation ${\boldsymbol\Gamma}$.\\

Returning to the general case we will show that any elliptic subspace  $S'\subset{\mathbb R}^N$ contains a completely
elliptic subspace $S\subset S'$ with the same set of states living on the two subspaces:
\begin{prop}\label{PropC}
Let $S'\subset{\mathbb R}^N$  be $M'$-elliptic. Then there exists a completely
elliptic subspace $S\subset S'$ with $\mbox{dim } S=M$ and ${\mathcal P}_{M',S'}={\mathcal P}_{M,S}$.
\end{prop}

According to this proposition we may confine ourselves to the case of a completely
elliptic subspace $S$. We want to analyze the set  ${\mathcal P}_{M,S}$.
Let us assume that a basis of $S$ is given and the $M$ basis vectors are written as the column vectors of an
$N\times M$-matrix $W$. $S$ being (completely) elliptic then entails the condition that some spin configuration
$({\mathbf s}_\mu)_{\mu=1,\ldots,N}$ can be obtained
by a linear combination of the $W_{\mu i}$:
\begin{equation}\label{DSG3}
{\mathbf s}_{\mu j}=\sum_{i=1}^{M}W_{\mu i}\,\Gamma_{i j},\;\mu=1,\ldots,N,\;j=1,\ldots,M
\;,
\end{equation}
or, in matrix notation,
\begin{equation}\label{DSG4}
 {\mathbf s}=W\,{\boldsymbol\Gamma}
 \;.
\end{equation}
The corresponding Gram matrix is $G={\mathbf s}\,{\mathbf s}^\top=W{\boldsymbol\Gamma}{\boldsymbol\Gamma}^\top W^\top$.
Then the condition that the ${\mathbf s}_\mu$ are unit vectors can be written as
\begin{equation}\label{DSG5}
1=G_{\mu\mu}=\left(W{\boldsymbol\Gamma}{\boldsymbol\Gamma}^\top W^\top\right)_{\mu\mu}
\;,\mu=1,\ldots,N
\;.
\end{equation}
With the definition $\Delta\equiv {\boldsymbol\Gamma}{\boldsymbol\Gamma}^\top\ge 0$ this condition assumes the form
\begin{equation}\label{DSG6}
 1=\left(W\,\Delta\,W^\top\right)_{\mu\mu}=\sum_{i,j=1}^{M} W_{\mu i}W_{\mu j}\Delta_{i j},\; \mu=1,\ldots,N
 \;,
\end{equation}
and can be considered as a system of $N$ inhomogeneous linear equations for the $\frac{1}{2}M(M+1)$ unknown entries $\Delta_{i j}$
of a symmetric $M\times M$ matrix. Its solution set will be an affine subspace of ${\mathbbm R}^{\frac{1}{2}M(M+1)}$,
where the latter space will be identified with ${\mathcal S}{\mathcal M}(M)$, the space of all real, symmetric $M\times M$ matrices.
The condition $\Delta\ge 0$ restricts the solution set of (\ref{DSG6})
to a compact convex subset of ${\mathbbm R}^{\frac{1}{2}M(M+1)}$
that is, by definition, non-empty for elliptic subspaces $S$.
We will refer to the system of equations (\ref{DSG6}) together with the condition that
$\Delta\ge 0$ as the ``additional degeneracy equation" (ADE).
Its set of solutions $\Delta\ge 0$ will be denoted by ${\mathcal S}_{ADE}$.
It can be shown that $G=W\,\Delta\,W^\top$ describes a $1:1$ correspondence between
the points of ${\mathcal S}_{ADE}$ and the Gram matrices of spin configurations living on $S$.

Consider an arbitrary solution $\Delta\in{\mathcal S}_{ADE}$.
Then there exists the square root ${\boldsymbol\gamma}$ such that
$\Delta={\boldsymbol\gamma}^2,\,{\boldsymbol\gamma}\ge 0$
and ${\mathbf s}\equiv W\,{\boldsymbol\gamma}$ will be a spin configuration living on $S$.
Any other spin configuration $\bar{\mathbf s}$ with the same Gram matrix $G=W\,\Delta\, W^\top$
must be of the form $\bar{\mathbf s}={\mathbf s}\,R$, with $R\in O(M)$, see Proposition \ref{Prop5}, and hence
\begin{equation}\label{DSGWDeltaR}
\bar{\mathbf s}=W\,{\boldsymbol\gamma}\,R= W\,\sqrt{\Delta}\,R
\;.
\end{equation}
The latter equation nicely captures the
separation of the degeneracy of ground states into rotational/reflectional degeneracy represented by $R$ and the additional
degeneracy represented by $\Delta$. This separation anticipates the result that the Lagrange parameters of the ground state
are unique, $\Lambda_0=\{\hat{\boldsymbol\lambda}\}$. Otherwise we would have a third kind of ``anomalous" degeneracy. But note that
the result $\Lambda_0=\{\hat{\boldsymbol\lambda}\}$ will only be proven in the sense of admitting $M$-dimensional ground states.
Insisting of the condition that $M\le3$ for physical ground states would open the possibility for anomalous degeneracy.\\

We will further investigate the degree of additional degeneracy.
According to the assumption of complete ellipticity there exists some ${\mathbf s}\in{\mathcal P}_{M,S}$ with
$\mbox{dim }{\mathbf s}=M$.  Such an ${\mathbf s}$ living on a completely elliptic
 subspace will be called a state of ``maximal dimension".
It follows that in the above representation ${\mathbf s}=W\,{\boldsymbol\Gamma}$ the matrix $\boldsymbol\Gamma$ must
have the rank $M$. Let $\Delta_0= {\boldsymbol\Gamma}\, {\boldsymbol\Gamma}^\top$, then also $\mbox{rank }\Delta_0=M$ which implies $\Delta_0> 0$.
The latter is equivalent to $\Delta_0$ lying in the interior of the convex set ${\mathcal S}_{ADE}$.

Now consider the homogeneous linear system of equations corresponding to (\ref{DSG6}):
\begin{equation}\label{DSG7}
0=\left(W\,\Delta\,W^\top\right)_{\mu\mu}=\sum_{i,j=1}^{M} W_{\mu i}W_{\mu j}\Delta_{i j}
=\mbox{Tr }\left( P_\mu\,\Delta\right),
\end{equation}
for all $ \mu=1,\ldots,N \;,$
where the rank $1$ matrices $P_\mu$ are defined by $(P_\mu)_{ij}\equiv W_{\mu i}W_{\mu j},\;i,j=1,\ldots,M$.
The $P_\mu$ are the projectors onto the $1$-dimensional subspaces spanned by the $\mu$-th row $W_\mu$ of $W$ multiplied
by $\| W_\mu\|^2$.

Recall that ${\mathcal S}{\mathcal M}(M)$ denotes the $M(M+1)/2$-dimensional space of all real, symmetric $M\times M$-matrices.
It will be equipped with the inner product $\langle A\,|\,B\rangle=\mbox{Tr } A B$ .
${\mathcal S}{\mathcal M}_+(M)\subset {\mathcal S}{\mathcal M}(M)$ denotes the closed, convex cone of
positively semi-definite matrices.
Further, let $P$ be  the subspace of ${\mathcal S}{\mathcal M}(M)$ spanned by the $P_\mu,\mu=1,\ldots,N,$
with dimension $\mbox{dim } P=p$. Then (\ref{DSG7})
says that $\Delta$ is lying in the orthogonal complement $P^\perp$ of $P$ in ${\mathcal S}{\mathcal M}(M)$.  Since the general solution
of (\ref{DSG6}) can be written as the sum of $\Delta_0$ and the general solution of (\ref{DSG7}) we have the following result:
\begin{prop}\label{PropADE}
With the preceding definitions, the set of solutions $\Delta\ge0$  of the ADE is the convex set
${\mathcal S}_{ADE}=\left(\Delta_0+P^\perp\right) \cap {\mathcal S}{\mathcal M}_+(M)$
and has the dimension $d\equiv M(M+1)/2\,-\,p= \mbox{dim } P^\perp$.
\end{prop}
According to this Proposition $d$ will be called the ``degree of additional degeneracy" or simply the ``degree" of the matrix $W$
the columns of which span an elliptic subspace $S$.
It vanishes, i.~e.~, $\Delta$ is unique iff the $P_\mu,\mu=1,\ldots,N$ span the total space ${\mathcal S}{\mathcal M}(M)$.
$p$ will be called the ``co-degree" of $W$.
We will also speak of the ``degree $d$ of ${\mathbf s}$" and the ``co-degree $p$ of ${\mathbf s}$" in the case of a state ${\mathbf s}$
of maximal dimension $M$ living on a completely elliptic subspace.

It can be shown that the co-degree is never smaller than the dimension:
\begin{prop}\label{ProppM}
$ M\le p \le N$.
\end{prop}

We close this subsection with two elementary examples.

\vspace{5mm}
\fbox{Example 3: The AF equilateral triangle ($N=3$)}

\vspace{5mm}
The AF equilateral spin triangle can be described by the Hamiltonian
\begin{equation}\label{ex21}
H=2({\mathbf s}_1\cdot{\mathbf s}_2+{\mathbf s}_2\cdot{\mathbf s}_3+{\mathbf s}_3\cdot{\mathbf s}_1)
 \;,
\end{equation}
and is the simplest example of a ``frustrated" spin system. This means that its ground state does not minimize each term of
(\ref{ex21}). This ground state is realized by any co-planar spin configuration with a mutual angle of $2\pi/3$ between any two
spin vectors. Hence it is essentially unique. We will use this well-known system to illustrate
the considerations of this subsection.

\begin{figure}[ht]
  \makebox[\linewidth][c]{\includegraphics[width=1.0\linewidth]{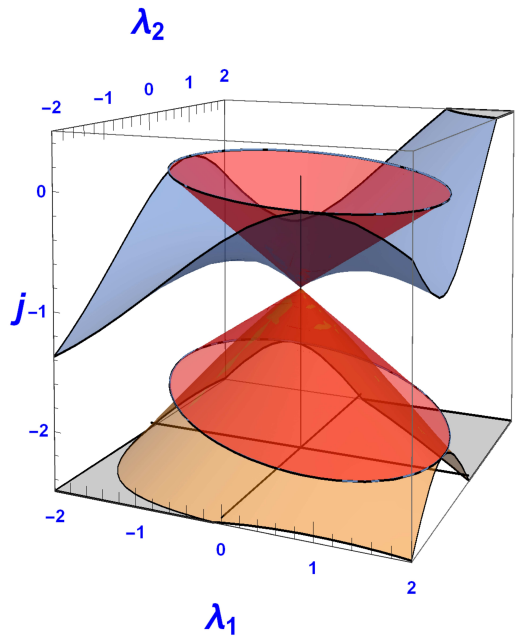}}%
  \caption[Example 3]
  {The two lowest eigenvalues $j_{min}(\boldsymbol{\lambda})$ and $j_{2}(\boldsymbol{\lambda})$
  of the dressed ${\mathbbm J}$-matrix for the AF equilateral triangle.
  $j_{min}(\boldsymbol{\lambda})$ has its maximum at the singular point ${\boldsymbol\lambda}={\mathbf 0}$
  where the Lagrange variety ${\mathcal V}$ can locally be approximated by a double cone (shown in red color).
 The coplanar ground state (\ref{ex210}) is living on the corresponding eigenspace of $({\mathbbm J}({\mathbf 0}),j_{min}({\mathbf 0}))$.}
  \label{FIGEX2}
\end{figure}

First we note that ${\mathbbm J}(\boldsymbol\lambda)$ assumes the form
\begin{equation}\label{ex22}
{\mathbbm J}(\boldsymbol\lambda)=
 \left(\begin{array}{ccc}
 \lambda_1 & 1 & 1 \\
 1 & \lambda_2 & 1 \\
 1 & 1 & -\lambda_1-\lambda_2 \\
\end{array}\right)
\;,
\end{equation}
which leads to the characteristic equation
\begin{eqnarray}\label{ex23}
0&=& \det\left({\mathbbm J}(\boldsymbol\lambda)-x\,{\mathbbm 1}\right)\\
\nonumber
&=&
 2 -\lambda_1 \lambda_2(\lambda_1 + \lambda_2)-x^3+x \left(\lambda_1^2+\lambda_1\lambda_2+\lambda_2^2+3\right)
   \;.\\
   \label{ex24}
\end{eqnarray}
It follows that $j_{min}({\boldsymbol\lambda})$ has its maximum $\hat{\jmath}$ at a singular point of the Lagrange variety
${\mathcal V}$
corresponding to ${\boldsymbol\lambda}={\mathbf 0}$ and the doubly degenerate eigenvalue
$\hat{\jmath}=j_{min}({\mathbf 0})=-1$, see Figure \ref{FIGEX2}.

A basis of the eigenspace of $({\mathbbm J}({\mathbf 0}),-1)$ is given by the two column vectors of
\begin{equation}\label{ex25}
W=\left(
\begin{array}{rr}
 -1 & -1 \\
 0 & 1 \\
 1 & 0 \\
\end{array}
\right)
\;.
\end{equation}
The solution of the corresponding ADE (\ref{DSG6}) is unique and given by
\begin{equation}\label{ex26}
 \Delta=\left(
\begin{array}{rr}
 1 & -\frac{1}{2} \\
 -\frac{1}{2} & 1 \\
\end{array}
\right)
\;.
\end{equation}
Its square root
\begin{equation}\label{ex27}
\sqrt{ \Delta}=
\left(
\begin{array}{rr}
 \frac{\sqrt{2+\sqrt{3}}}{2} & \frac{1}{4}
   \left(\sqrt{2}-\sqrt{6}\right) \\
 \frac{1}{4} \left(\sqrt{2}-\sqrt{6}\right) &
   \frac{\sqrt{2+\sqrt{3}}}{2} \\
\end{array}
\right)
\end{equation}
leads to
\begin{equation}\label{ex28}
{\mathbf s}=W\,\sqrt{ \Delta}=
\left(
\begin{array}{rr}
 -\frac{1}{\sqrt{2}} & -\frac{1}{\sqrt{2}} \\
 \frac{1}{4} \left(\sqrt{2}-\sqrt{6}\right) &
   \frac{\sqrt{2+\sqrt{3}}}{2} \\
 \frac{\sqrt{2+\sqrt{3}}}{2} & \frac{1}{4}
   \left(\sqrt{2}-\sqrt{6}\right) \\
\end{array}
\right)\;.
\end{equation}
This is indeed a ground state of (\ref{ex21}) albeit in an unusual form. To obtain a more familiar representation
we multiply (\ref{ex28}) with the rotation matrix
\begin{equation}\label{ex29}
 R=
 \left(
\begin{array}{rr}
 -\frac{1}{\sqrt{2}} & \frac{1}{\sqrt{2}} \\
 -\frac{1}{\sqrt{2}} & -\frac{1}{\sqrt{2}} \\
\end{array}
\right)
\end{equation}
and obtain
\begin{equation}\label{ex210}
\bar{\mathbf s}=W\,\sqrt{ \Delta}\,R=
 \left(
\begin{array}{rr}
 1 & 0 \\
 -\frac{1}{2} & -\frac{\sqrt{3}}{2} \\
 -\frac{1}{2} & \frac{\sqrt{3}}{2} \\
\end{array}
\right)
\;.
\end{equation}

The preceding example illustrates the construction of ground states from an elliptic eigenspace,
but it does not show any additional degeneracy since $d=0$. Hence we will provide another example where additional
degeneracy occurs.

\vspace{5mm}

\fbox{Example 4: The AF bow tie ($N=5$)}

\vspace{5mm}

\begin{figure}[ht]
  \centering
    \includegraphics[width=1.0\linewidth]{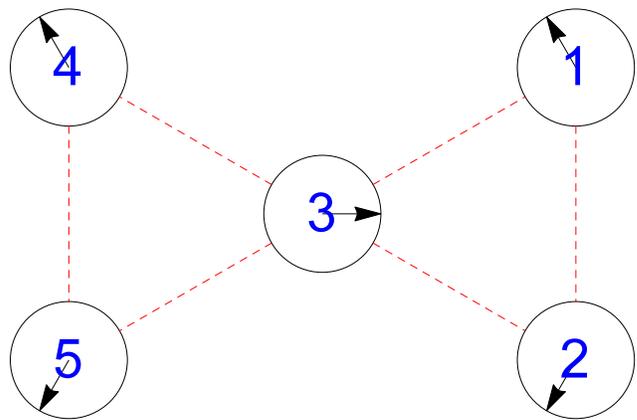}
  \caption[Example 3]
  {The AF  bow tie and a co-planar ground state indicated by arrows with a mutual angle of $2\pi/3$ between
  neighboring spins. }
  \label{FIGEX3}
\end{figure}

The AF ``bow tie" consists of two corner-sharing triangles, see Figure \ref{FIGEX3}, and can be described by the Hamiltonian
\begin{equation}\label{ex31}
H=2({\mathbf s}_1\cdot{\mathbf s}_2+{\mathbf s}_1\cdot{\mathbf s}_3+{\mathbf s}_2\cdot{\mathbf s}_3
+{\mathbf s}_3\cdot{\mathbf s}_4+{\mathbf s}_3\cdot{\mathbf s}_5+{\mathbf s}_4\cdot{\mathbf s}_5)
\;,
\end{equation}
that can be viewed as the sum of two triangle Hamiltonians $H_1,\;H_2$ of the kind (\ref{ex21}) considered in Example 3.
It is possible to minimize $H_1$ and $H_2$ simultaneously, for example by the co-planar ground state indicated
in Figure \ref{FIGEX3}. Moreover, one can rotate the spins with number $1$ and $2$ about the axis of the central spin
with number $3$ independently of the remaining spins with number $4$ and $5$. This yields a one-parameter family
of ground states that are not $O(3)$-equivalent and hence an example of additional degeneracy of degree $1$.

It remains to show how these facts about the bow tie's ground states are reproduced by the present theory.
First consider the dressed ${\mathbbm J}$-matrix of the form
\begin{equation}\label{ex32}
{\mathbbm J}({\boldsymbol\lambda})=
\left(
\begin{array}{ccccc}
 \lambda_1 & 1 & 1 & 0 & 0 \\
 1 & \lambda_2 & 1 & 0 & 0 \\
 1 & 1 & \lambda_3 & 1 & 1 \\
 0 & 0 & 1 &\lambda_4 & 1 \\
 0 & 0 & 1 & 1 & -\lambda_1-\lambda_2-\lambda_3-\lambda_4
 \\
\end{array}
\right)
\;.
\end{equation}
We have to find a ${\boldsymbol\lambda}\in\Lambda$ such that $j_{min}({\boldsymbol\lambda})$ assumes its maximum.
The present theory does not provide a silver bullet to fulfill this task in general and we do not want to
anticipate the results of subsection \ref{sec:DF} concerning the ground state gauge for fused spin systems.
One possibility to tackle the problem would be to find any ground state by whatever means (numerical or analytical)
and to calculate its Lagrange parameters according to (\ref{D7}). Sometimes it will be possible
to estimate the exact values from its numerical approximations. In our case we simply take the co-planar ground state
indicated in Figure \ref{FIGEX3} and obtain the corresponding   ${\boldsymbol\lambda}$ as
\begin{equation}\label{ex33}
\lambda_3=\frac{4}{5},\;
\lambda_1=\lambda_2=\lambda_4=\lambda_5=-\frac{1}{5}\;.
\end{equation}
This leads to the maximal eigenvalue $\hat{\jmath}=j_{min}({\boldsymbol\lambda})=-\frac{6}{5}$.
It turns out that for these values the eigenspace of
$({\mathbbm J}({\boldsymbol\lambda}),\hat{\jmath})$ has the dimension $M=3$.
A basis of it is given by the column vectors of the following matrix
\begin{equation}\label{ex34}
W=\left(
\begin{array}{rrr}
 1 & 1 & -1 \\
 0 & 0 & 1 \\
 -1 & -1 & 0 \\
 0 & 1 & 0 \\
 1 & 0 & 0 \\
\end{array}
\right)\;.
\end{equation}

The rank $1$ matrices $P_\mu,\;\mu=1,\ldots,5$ generated by the rows of $W$ span a $5$-dimensional subspace of ${\mathcal S}{\mathcal M}(3)$.
Hence Proposition \ref{PropADE} yields an additional degeneracy of degree
\begin{equation}\label{ex36}
d=\frac{M(M+1)}{2} -p=\frac{3\times 4}{2}-5=1
\;.
\end{equation}

In accordance with this the ADE (\ref{DSG6}) has a one-parameter family $\Delta(\delta)$ of solutions
\begin{equation}\label{ex35}
 \Delta(\delta)
 =
\left(
\begin{array}{rrr}
 1 & -\frac{1}{2} & \delta  \\
 -\frac{1}{2} & 1 & \frac{1}{2}-\delta  \\
 \delta  & \frac{1}{2}-\delta  & 1 \\
\end{array}
\right)\;.
\end{equation}

\begin{figure}[ht]
  \centering
    \includegraphics[width=1.0\linewidth]{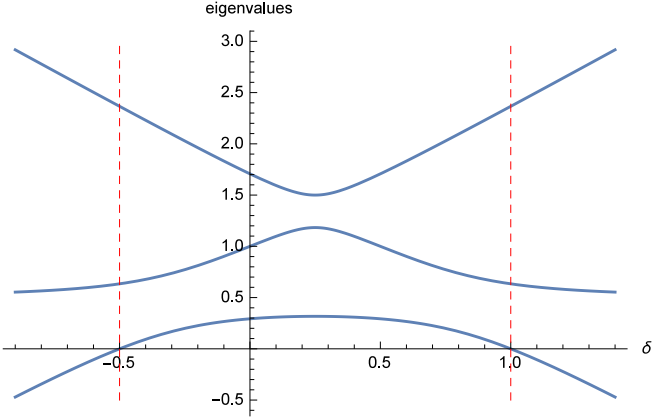}
  \caption[Example 3]
  {For the AF bow tie the eigenvalues of $\Delta(\delta)$ are non-negative for $-\frac{1}{2}\le \delta\le 1$. }
  \label{FIGEX3E}
\end{figure}

The eigenvalues of $\Delta(\delta)$ are shown in Figure \ref{FIGEX3E}. It follows by inspection, and can easily be derived analytically,
that $\Delta(\delta)\ge 0$ for $-1/2\le \delta \le 1$. For $-1/2<\delta < 1$, $\Delta(\delta)$ represents a one-parameter
family of $3$-dimensional ground states, whereas at the endpoints of the interval $[-1/2,1]$ the rank of  $\Delta(\delta)$ and hence
the dimension of the corresponding ground states is reduced to $2$. This complies with the geometric picture of additional degeneracy
of the bow tie's ground states sketched above.

To further confirm the accordance between the geometric picture and the theory's results we give the result for the Gram matrix
$G(\delta)=W\, \Delta(\delta)\,W^\top$ of the considered one-parameter family:
\begin{equation}\label{ex35}
 G(\delta)=\left(
\begin{array}{rrrrr}
 1 & -\frac{1}{2} & -\frac{1}{2} & \delta  & \frac{1}{2}-\delta
   \\
 -\frac{1}{2} & 1 & -\frac{1}{2} & \frac{1}{2}-\delta  & \delta
   \\
 -\frac{1}{2} & -\frac{1}{2} & 1 & -\frac{1}{2} & -\frac{1}{2} \\
 \delta  & \frac{1}{2}-\delta  & -\frac{1}{2} & 1 & -\frac{1}{2}
   \\
 \frac{1}{2}-\delta  & \delta  & -\frac{1}{2} & -\frac{1}{2} & 1
   \\
\end{array}
\right)
\;.
\end{equation}
Recall the $G_{\mu\nu}={\mathbf s}_\mu\cdot{\mathbf s}_\nu,\;\mu,\nu=1,\ldots,5$.
One observes that the mutual scalar products are constant within the triangles $(1,2,3)$ and $(3,4,5)$ and assume the value $\cos 2\pi/3=-1/2$
corresponding to the triangle's ground state considered in Example 2. Only the scalar products between the two groups $(1,2)$
and $(4,5)$ vary with $\delta$ as it must be if the corresponding spins are independently rotated.

\subsection{Fusion}\label{sec:DF}

This subsection contains some results on a generalization of Example 4 in connection with the Lagrange variety approach.
It illustrates some aspects of this approach but will not be presupposed in the following sections.

The bow tie example is an instance of the general process of ``fusing" two spin systems. By this we mean the
union of two spin systems that are disjoint except for a single spin. In the Example 4
we may consider two triangles with spin numbers $(1,2,3)$ and $(3,4,5)$ with the common spin number $3$. The bow tie then results
from the union $(1,2,3,4,5)$, see Figure \ref{FIGEX3}.

Returning to the general case we denote by $\Sigma_1=(1,\ldots,N_1)$ and $\Sigma_2=(N_1,\ldots,N_1+N_2-1)$ two
sets of spin numbers that are disjoint except for the common spin with number
$N_1$ and by $\Sigma=(1,\ldots,N_1,N_1+1,\ldots,N)$ their fusion, where $N\equiv N_1+N_2-1$.
The corresponding Hamiltonians are
\begin{eqnarray}\label{DF1}
 H_1 &=& \sum_{\mu,\nu=1}^{N_1}\,J_{\mu\nu}^{(1)}\, {\mathbf s}_\mu\cdot  {\mathbf s}_\nu , \\
 \label{DF2}
 H_2 &=& \sum_{\mu,\nu=N_1}^N\,J_{\mu\nu}^{(2)}\, {\mathbf s}_\mu\cdot  {\mathbf s}_\nu , \\
 \label{DF3}
 H &=& \sum_{\mu,\nu=1}^N\,J_{\mu\nu}\, {\mathbf s}_\mu\cdot  {\mathbf s}_\nu , \\
 \nonumber
 \mbox{where}&&\\
 \label{DF4}
 J_{\mu\nu}&=&\left\{\begin{array}{r@{\quad:\quad}l}
 J_{\mu\nu}^{(1)}& 1\le \mu,\nu\le N_1,\\
 J_{\mu\nu}^{(2)}& N_1\le \mu,\nu\le N,\\
 0& \mbox{otherwise}.
 \end{array} \right.
\end{eqnarray}
We will also speak of the ``large spin system", corresponding to $\Sigma$ and of the two ``subsystems", corresponding to
$\Sigma_1$ and $\Sigma_2$, without danger of misunderstanding.
Let ${\mathbf s}_\mu^{(1)},\,\mu=1,\ldots,N_1,$ and ${\mathbf s}_\mu^{(2)},\,\mu=N_1,\ldots,N,$ be states of the two subsystems.
A usual, we consider the ${\mathbf s}^{(i)}$ as $N_i\times M_i$-matrices.
Let ${\mathbf S}^{(i)}$ be the $N\times (M_1+M_2)$-matrices obtained by copying the ${\mathbf s}^{(i)}$ into the larger matrix and padding the
remaining entries by zeroes such that all rows of ${\mathbf S}^{(1)}$ are orthogonal to all rows of ${\mathbf S}^{(2)}$:
\begin{eqnarray} \nonumber
  {\mathbf S}^{(1)}_{\mu,i} &\equiv& \left\{\begin{array}{l@{\;:\;}l}
{\mathbf s}^{(1)}_{\mu,i}& 1\le \mu\le N_1\mbox{ and }1\le i \le M_1,\\
0& \mbox{otherwise},
 \end{array} \right. \\
 \label{DF4a}  &&\\
 \nonumber
 {\mathbf S}^{(2)}_{\mu,i} &\equiv& \left\{\begin{array}{l@{\;:\;}l}
{\mathbf s}^{(2)}_{\mu,i-M_1}&N_1\le \mu\le N \mbox{ and }M_1 < i \le M_1+M_2,\\
0& \mbox{otherwise}.
 \end{array} \right. \\
 \label{DF4b} &&
\end{eqnarray}

Then there exists an $R\in O(M_1+M_2)$ such that
\begin{equation}\label{DF5}
 R\,{\mathbf S}_{N_1}^{(2)}={\mathbf S}_{N_1}^{(1)}
 \;.
\end{equation}
We set
\begin{equation}\label{DF6}
 \bar{\mathbf S}_\nu^{(2)} \equiv R\, {\mathbf S}_\nu^{(2)},\; \nu=N_1,\ldots,N
 \;,
\end{equation}
and
\begin{equation}\label{DF7}
 {\mathbf s}_\mu \equiv
 \left\{\begin{array}{r@{\quad:\quad}l}
{\mathbf S}_\mu^{(1)}& 1\le \mu\le N_1,\\
\bar{\mathbf S}_\mu^{(2)}&  N_1\le \mu\le N,
 \end{array} \right.
\end{equation}
for all $\mu=1,\ldots N$. Obviously, ${\mathbf s}$ is a state of the large spin system
that will be called the ``fusion" of the states ${\mathbf s}^{(1)}$ and ${\mathbf s}^{(1)}$.
The fusion of two  states is not unique since there are many rotations/reflections $R\in O(M_1+M_2)$ satisfying (\ref{DF5}).
Recall that this non-uniqueness leads to the additional degeneracy in the bow tie Example 4.
We have the following results:
\begin{prop}\label{PropF1}
Under the preceding definitions the following holds:
\begin{itemize}
     \item[(i)]If ${\mathbf s}^{(1)}$ and ${\mathbf s}^{(1)}$ are ground states of $H_1$ and $H_2$, resp.~, and
     ${\mathbf s}$ is a fusion of ${\mathbf s}^{(1)}$ and ${\mathbf s}^{(2)}$, then ${\mathbf s}$ will be a ground state of $H$.
   \item[(ii)] Every ground state ${\mathbf s}$ of $H$ can be obtained by a fusion of two ground states ${\mathbf s}^{(1)}$ and ${\mathbf s}^{(2)}$
   of $H_1$ and $H_2$, resp.~.
  \end{itemize}
\end{prop}

\begin{prop}\label{PropF2}
Let the ground states ${\mathbf s}^{(i)}$ of $H_i$ be of maximal dimension $M_i$ for $i=1,2$.
Then there exists a fusion ${\mathbf s}$ of ${\mathbf s}^{(1)}$ and ${\mathbf s}^{(2)}$ that is a ground state of maximal
dimension $M$ of $H$. Let $d$ denote its degree and $p$ its co-degree
and analogously $d_i$ the degree and $p_i$ the co-degree of ${\mathbf s}^{(i)}$ for $i=1,2$.
Then the following holds:
 \begin{itemize}
     \item[(i)]$M= M_1+M_2-1$,
   \item[(ii)] $p=p_1+p_2-1$,
   \item[(iii)] $d=d_1+d_2+(M_1-1)(M_2-1)$.
  \end{itemize}
\end{prop}
In the bow tie example 4 we have indeed  $M=2+2-1=3, \; p= 3+3-1=5$ and $d=0+0+1\times 1 =1$.

Of course, the fusion process can be iterated and yields some kind of tree-like spin structures.
But not every system of corner-sharing triangles can be obtained by iterative fusions of triangles, e.~g., the cuboctahedron.

We want to show in more details how the fusion process complies with the Lagrange variety approach. First, it will be obvious
how to define the fusion of the corresponding ${\mathbbm J}$-matrices, ${\mathbbm J}^{(1)}$ and ${\mathbbm J}^{(2)}$
such that ${\mathbbm J}$ contains ${\mathbbm J}^{(1)}$ and ${\mathbbm J}^{(2)}$ as sub-matrices. From this it follows
that the Lagrange parameters ${\boldsymbol\kappa}$ occurring in (\ref{D7}) will be additive,
\begin{equation}\label{DF8}
 {\boldsymbol\kappa}={\boldsymbol\kappa}^{(1)}+{\boldsymbol\kappa}^{(2)}
 \;,
\end{equation}
taking into account the embedding of the two sets of spin numbers $\Sigma_1$ and $\Sigma_2$ into $\Sigma=\{1,\ldots,N\}$. Consequently,
\begin{eqnarray}\label{DF9}
  \bar{\kappa} &=& \frac{1}{N}\left( N_1\,\bar{\kappa}^{(1)}  + N_2\,\bar{\kappa}^{(2)}   \right), \\
  \label{DF10}
  \lambda_\mu &=&\kappa_\mu^{(1)}+\kappa_\mu^{(2)}-\bar{\kappa}, \\
  \label{DF11}
  E_{min} &=& -N\,\bar{\kappa}=-N_1\,\bar{\kappa}^{(1)}-N_2\,\bar{\kappa}^{(2)}\\
  \label{DF12}
  &=& E_{min}^{(1)}+E_{min}^{(2)}.
\end{eqnarray}
The latter equation is also obvious from the equation $H=H_1+H_2$ and the possibility to minimize each term independently.

Equation (\ref{DF10}) implies that the ground state gauge parameters ${\lambda}_\mu$ will not be additive, i.~e.~,
${\lambda}_\mu\neq {\lambda}_\mu^{(1)}+{\lambda}_\mu^{(2)}$. This will be illustrated
by reconsidering the bow tie example 4. Here we have
\begin{eqnarray} \label{DF13a}
   && \kappa_1^{(1)}=\kappa_2^{(1)}=\kappa_3^{(1)}=1,\; \overline{\kappa}^{(1)}=1,\; \lambda_\mu^{(1)}=0,\\
    \label{DF13b}
   && \kappa_3^{(2)}=\kappa_4^{(2)}=\kappa_5^{(2)}=1,\; \overline{\kappa}^{(2)}=1,\; \lambda_\mu^{(2)}=0,\\
    \label{DF13c}
   && \kappa_1=\kappa_2=\kappa_4=\kappa_5=1,\, \kappa_3=2,\\
    \label{DF13d}
   && \bar{\kappa}\stackrel{(\ref{DF9})}{=}\frac{1}{5}
   \left( 3\times 1+3\times 1\right) =\frac{6}{5},
   \\  \label{DF13e}
   && \lambda_1=\lambda_2=\lambda_4=\lambda_5=1-\frac{6}{5}=-\frac{1}{5},\\
    \label{DF13f}
   &&   \lambda_3=1+1-\frac{6}{5}=\frac{4}{5},
\end{eqnarray}
in accordance with (\ref{ex33}).

\section{Elliptic points of the Lagrange variety}\label{sec:SP}
This section is rather technical in character but it is crucial for the following section \ref{sec:EU} on existence and uniqueness of ground states.

We again consider an $N\times M$-matrix ${\mathbf s}$ with $N$ row vectors ${\mathbf s}_\mu$ and $M$ column vectors ${\mathbf s}_i$
and reconsider the SSE (\ref{D7}) written in the form of an eigenvalue equation
\begin{eqnarray}\label{SP1}
{\mathbbm J}(\boldsymbol\lambda)\, {\mathbf s}_i &=&-\bar{\kappa}\,{\mathbf s}_i,\mbox{ for }i=1,\ldots,M,\\
\nonumber
\mbox{where}&&\\
\label{SP2}
{\mathbf s}_\mu\cdot{\mathbf s}_\mu&=&1, \mbox{ for } \mu=1,\ldots,N
\;.
\end{eqnarray}
Hence to each solution of (\ref{SP1}),(\ref{SP2})  there belongs a point
$({\boldsymbol\lambda},-\bar{\kappa})$ of the Lagrange variety ${\mathcal V}$, see (\ref{DdefV}), such that the
eigenspace of $({\mathbbm J}(\boldsymbol\lambda),-\bar{\kappa})$ is elliptic and vice versa. We will call such points of ${\mathcal V}$
``elliptic". It is the aim of the present section to closer characterize elliptic points of ${\mathcal V}$.

Let the symbol ${\mathbf D}$ denote
the vector ${\mathbf D}=(D_1,D_2\ldots,D_N)$ of diagonal matrices defined by
\begin{equation}\label{EG2}
 \left(D_\mu\right)_{ij}\equiv \delta_{\mu i}\delta_{ij},\mbox{ for all } \mu,i,j =1,\ldots,N
 \;.
\end{equation}

Note that
\begin{equation}\label{EG2a}
 {\mathbbm J}({\boldsymbol\lambda})={\mathbbm J}({\mathbf 0})+  {\mathbf D}\cdot {\boldsymbol\lambda}
 \;,
\end{equation}
which will be used below in the application of perturbation theory.

Now we consider a general point $({\boldsymbol\lambda},x)\in {\mathcal V}$ and the eigenspace $S$ of $({\mathbbm J}(\boldsymbol\lambda),x)$.
Let $S_1\equiv \{\varphi \in S |\,\|\varphi\|=1\}$. It follows that for all $\varphi\in S_1$ the function
\begin{eqnarray}\nonumber
  h_\varphi &:& \Lambda\longrightarrow {\mathbbm R} \\
 \label{SP3}
 h_\varphi({\boldsymbol\mu}) &\equiv& \left\langle \varphi \left| {\mathbf D}\cdot {\boldsymbol\mu}\right| \varphi\right\rangle
\end{eqnarray}
will be linear. Hence its graph will be a hyperplane of $\Lambda\times {\mathbbm R}$ containing the origin $({\mathbf 0},0)$.
Further it follows that the ``super-graph" of $h_\varphi$,
\begin{equation}\label{SP4}
 H_\varphi^+ \equiv \{({\boldsymbol\mu},y)\in \Lambda\times {\mathbbm R}\left|
 y\ge \left\langle \varphi \left| {\mathbf D}\cdot {\boldsymbol\mu}\right| \varphi\right\rangle\right. \}
\end{equation}
will be an upper closed half-space of $\Lambda\times {\mathbbm R}$. Analogously, the ``sub-graph" of  $h_\varphi$,
\begin{equation}\label{SP5}
 H_\varphi^- \equiv \{({\boldsymbol\mu},y)\in \Lambda\times {\mathbbm R}\left|
 y\le \left\langle \varphi \left| {\mathbf D}\cdot {\boldsymbol\mu}\right| \varphi\right\rangle\right. \}
\end{equation}
will be a lower closed half-space of $\Lambda\times {\mathbbm R}$ such that\\
$H_\varphi^-= - \,H_\varphi^+$.\\

Next we define the upper cone ${\mathcal C}^+({\boldsymbol\lambda},x)$ and the lower cone
${\mathcal C}^-({\boldsymbol\lambda},x)$ by
\begin{eqnarray}\label{SP8}
  {\mathcal C}^+({\boldsymbol\lambda},x) &\equiv& \bigcap_{\varphi\in S_1}H^+_\varphi, \\
  \label{SP9}
  {\mathcal C}^-({\boldsymbol\lambda},x) &\equiv& \bigcap_{\varphi\in S_1}H^-_\varphi= -\,  {\mathcal C}^+({\boldsymbol\lambda},x)
   \;.
\end{eqnarray}
Both cones are closed convex cones in the sense of \cite{R97}.
It may be helpful to appeal to the analogy with the forward and backward light cone in special relativity, but
note that the above-defined cones will not be elliptic ones except for special cases as given by Example 3, see Figure \ref{FIGEX2}.

We thus have attached to each point $({\boldsymbol\lambda},x)$ of the Lagrange variety ${\mathcal V}$ two cones
${\mathcal C}^+({\boldsymbol\lambda},x)$ and  ${\mathcal C}^-({\boldsymbol\lambda},x)$. Recall that
at a regular point $({\boldsymbol\lambda},x)$ of
${\mathcal V}$ the eigenspace $S$ of $({\boldsymbol\lambda},x)$ will be one-dimensional,
see  Proposition \ref{PropV}, hence there is only one function
$h_\varphi, \; \varphi \in S_1$, since $h_\varphi=h_{-\varphi}$. It follows that $ {\mathcal C}^+({\boldsymbol\lambda},x)= H^+_\varphi$,
i.~e.~, the upper cone degenerates to an upper closed half-space, analogously for ${\mathcal C}^-({\boldsymbol\lambda},x)= H^-_\varphi$.
In contrast to this, the degenerate points of ${\mathcal V}$ will always have proper cones.

\begin{defi}\label{defvert}
  Let $({\boldsymbol\lambda},x)$ be a point of the Lagrange variety ${\mathcal V}$.
  The upper cone ${\mathcal C}^+({\boldsymbol\lambda},x)$ will be called ``vertical" iff it
  is contained in the upper closed half-space $H^+$,
  \begin{equation}\label{SP10a}
{\mathcal C}^+({\boldsymbol\lambda},x)\subset
H^+\equiv \{({\boldsymbol\mu},y)|{\boldsymbol\mu}\in\Lambda \mbox{ and } y\ge 0\}
\;,
\end{equation}
  This is, of course, equivalent to the statement that
\begin{equation}\label{SP1b0}
{\mathcal C}^-({\boldsymbol\lambda},x)\subset
H^-\equiv \{({\boldsymbol\mu},y)|{\boldsymbol\mu}\in\Lambda \mbox{ and } y\le 0\}
\;,
\end{equation}
and hence also in this case the lower cone will be called ``vertical".
Without danger of confusion we will also say that the point $({\boldsymbol\lambda},x)$  of  ${\mathcal V}$
is ``vertical" iff one of the above conditions is satisfied.
\end{defi}

\begin{figure}[ht]
  \centering
    \includegraphics[width=1.0\linewidth]{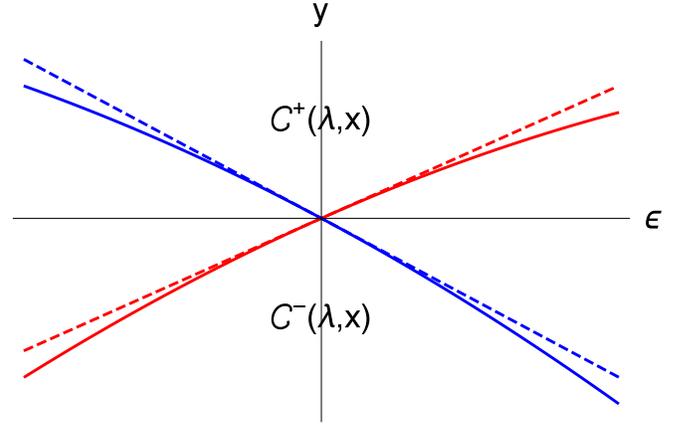}
  \caption[Example 1]
  {Schematic representation of the intersection of the upper/lower cone ${\mathcal C}^\pm ({\boldsymbol\lambda},x)$
  with some plane $L_0\times {\mathbbm R}$. It is bounded by the two dashed lines
  $y=\epsilon\, a_{min}$ and $y=\epsilon\, a_{max}$ obtained by perturbation theory, see (\ref{SP12}) and (\ref{SP13}).
  The eigenvalues $x_i(\epsilon),\;i=1,\ldots,n,$ are given by the continuous (red and blue) curves, where we have chosen $n=2$
  for the sake of simplicity. Note also that the shown intersection is typical for vertical cones, see (\ref{SP14}).
  }
  \label{FIGC}
\end{figure}

It will be in order to closer examine the geometrical meaning of the upper (lower) cones. To this end
consider $({\boldsymbol\lambda},x)\in{\mathcal V}$ and $S$ being the eigenspace of $({\mathbbm J}({\boldsymbol\lambda}),x)$,
such that $n\equiv \dim S >1$. Let ${\mathbbm Q}$ denote the projector onto $S$.
We fix some ${\boldsymbol\mu}\in\Lambda,\;{\boldsymbol\mu}\cdot{\boldsymbol\lambda}=0$ and define
$L_0\equiv \{\alpha\,{\boldsymbol\mu}|\alpha\in{\mathbbm R}\}$.
Further consider the eigenvalues $x_i(\epsilon),\;i=1,\ldots,n$ of ${\mathbbm J}({\boldsymbol\lambda}+\epsilon{\boldsymbol\mu})$.
These eigenvalues are obtained in the order ${\mathcal O}(\epsilon)$ by first order degenerate perturbation theory that is usually treated in
textbooks on quantum theory. For a mathematically rigorous account, see, e.~g., \cite{LT85}, chapter 11.
According to this theory the eigenvalues $x_i(\epsilon),\;i=1,\ldots, n$ in a neighbourhood of $\epsilon=0$ analytically depend on $\epsilon$
and satisfy
\begin{equation}\label{SP11}
 x_i(\epsilon)= x+ \epsilon \langle \varphi_i | {\mathbf D}\cdot {\boldsymbol\mu}|\varphi_i\rangle+{\mathcal O}(\varepsilon^2)
 \;,
\end{equation}
where $(\varphi_i)_{i=1,\ldots,n}$ is some eigenbasis of ${\mathbbm Q}\,{\mathbf D}\cdot {\boldsymbol\mu}\,{\mathbbm Q}$.
Of course, the eigenvalues of ${\mathbbm Q}\,{\mathbf D}\cdot {\boldsymbol\mu}\,{\mathbbm Q}$ may still be partially
degenerate and accordingly the eigenbasis may not be unique. Geometrically speaking, the tangents to the curves
$\epsilon\mapsto x_i(\epsilon)$ at $(0,x)$ have the slope $a_i\equiv \langle \varphi_i | {\mathbf D}\cdot {\boldsymbol\mu}|\varphi_i\rangle$.
Let $a_{min}=\mbox{Min}\{a_i | i=1,\ldots,n\}$ and $a_{max}=\mbox{Max}\{a_i | i=1,\ldots,n\}$ denote the
extremal slopes. These are connected to the upper and lower cone as follows, see also Figure \ref{FIGC}:
\begin{eqnarray}\nonumber
{\mathcal C}^+
({\boldsymbol\lambda},x)\cap \left(L_0\times {\mathbbm R}\right) &=&
\{ (\epsilon{\boldsymbol\mu},y)\in\left(L_0\times {\mathbbm R}\right)|y\ge\epsilon\,  a_{max}\\
\label{SP12}
&& \mbox{ and }y\ge\epsilon\,  a_{min} \}
\;.
\end{eqnarray}
Analogously,
\begin{eqnarray}\nonumber
{\mathcal C}^-
({\boldsymbol\lambda},x)\cap \left(L_0\times {\mathbbm R}\right) &=&
\{ (\epsilon{\boldsymbol\mu},y)\in\left(L_0\times {\mathbbm R}\right)|y\le\epsilon\,  a_{max}\\
\label{SP13}
&& \mbox{ and }y\le\epsilon\,  a_{min} \}
\;.
\end{eqnarray}
Especially, ${\mathcal C}^\pm ({\boldsymbol\lambda},x)$ is vertical iff for all
${\boldsymbol\mu}\in\Lambda$ such that ${\boldsymbol\mu}\cdot{\boldsymbol\lambda}=0$
we have
\begin{equation}\label{SP14}
 a_{min}\,\le 0\, \le a_{max}
 \;.
\end{equation}

Now we can formulate the main result of this section.
\begin{theorem}\label{Theorem1}
A point $({\boldsymbol\lambda},x)$ of the Lagrange variety ${\mathcal V}$ is elliptic iff it is vertical.
\end{theorem}
 The main application of this theorem will be given in section \ref{sec:EU} where we consider the case that $j_{min}({\boldsymbol\lambda})$
 assumes its maximum $\hat{\jmath}$ at some  ${\boldsymbol\lambda}\in\widehat{J}$. Then it follows that $({\boldsymbol\lambda},\hat{\jmath})\in{\mathcal V}$
 is vertical and hence Theorem \ref{Theorem1} assures the existence of a ground state ${\mathbf s}$
 that lives on the eigenspace of $({\mathbbm J}({\boldsymbol\lambda}),\hat{\jmath})$.
 However, if $\mbox{dim}({\mathbf s})>3$ for all such ground states we have to look
 for other solutions of the SSE in order to find physical ground states,
 but in this case Theorem \ref{Theorem1} is still helpful since it says that we
 only have to look at vertical points of ${\mathcal V}$.

\section{Existence and uniqueness of ground states}\label{sec:EU}

The headline of this section must not be understood literally, since the existence of ground states is almost trivial
and they are not unique already due to rotational/reflectional degeneracy. What we rather mean is that (1)
there exists a ground state ${\mathbf s}$ living in the eigenspace $S$ of $({\mathbbm J}({\boldsymbol\lambda}),j_{min}({\boldsymbol\lambda}))$
for all ${\boldsymbol\lambda}\in\widehat{J}\subset \Lambda$
and (2) that $\widehat{J}$ consists of a single point, $\widehat{J}=\{\hat{{\boldsymbol\lambda}}\}$.
Recall that according to Proposition \ref{Prop2} the function $j_{min}({\boldsymbol\lambda})$ assumes its maximum $\hat{\jmath}$
at some compact, convex set $\widehat{J}\subset \Lambda$.
The price that we have to pay for proving these results is that $\mbox {dim }({\mathbf s})$ may be larger than $3$ for all
${\mathbf s}$ living on $S$ and that one has to look for other elliptic/vertical points of ${\mathcal V}$ in order to find
physical ground states.

We then state the first result:
\begin{theorem}\label{Theorem2}
All points $({\boldsymbol\lambda},\hat{\jmath})\in {\mathcal V}$ are elliptic for ${\boldsymbol\lambda}\in\widehat{J}$.
\end{theorem}
For the proof it suffices to note that (\ref{SP14}) is necessary in order that $j_{min}({\boldsymbol\lambda})$ assumes its maximum
at ${\boldsymbol\lambda}\in\widehat{J}$. Hence $({\boldsymbol\lambda},\hat{\jmath})$ is vertical and, by Theorem \ref{Theorem1},
also elliptic, i.~e.~, there exists a ground state  ${\mathbf s}$ living on $S$. Hence the set $\Lambda_0$ introduced after (\ref{D10})
is shown to be identical with $\widehat{J}$.

The second result of this section is
\begin{theorem}\label{Theorem3}
$\widehat{J}$ consists of a single point,
$\widehat{J}=\{\hat{{\boldsymbol\lambda}}\}$.
\end{theorem}

We have already pointed out that Theorem \ref{Theorem3} in a sense restricts the degeneracy of ground states to rotational/reflectional
degeneracy and additional degeneracy as defined in section \ref{sec:DAD}. Here we will explain some consequences for symmetric
spin systems although a systematic account of these is beyond the realm of the present paper. Let $\Pi\in O(N)$ denote the linear representation
of some permutation $\pi\in{\mathcal S}_N$ generated by accordingly permuting the standard basis of ${\mathbbm R}^N$
and ${\sf S}_N$ be the group of such $\Pi$. Let ${\sf Gr}$ be the group of ``symmetries" of a given spin system defined by
\begin{equation}\label{EU1}
{\sf Gr}\equiv\{\Pi\in{\sf S}_N\; |\;\Pi\,{\mathbbm J}={\mathbbm J}\,\Pi\}
\;.
\end{equation}
The corresponding subgroup of ${\mathcal S}_N$ will be denoted by ${\mathcal Gr}$.
It follows that $\Pi\in{\sf Gr}$ operates on $\Lambda$ via $ \Pi\,{\mathbbm J}({\boldsymbol\lambda})\Pi^{-1} = {\mathbbm J}({\boldsymbol\lambda}')$
where ${\boldsymbol\lambda}'_\mu={\boldsymbol\lambda}_{\pi^{-1}(\mu)},\;\mu=1,\ldots, N$.
Moreover, $\widehat{J}$ will be invariant under this action and hence, by Theorem \ref{Theorem3}, $\hat{\boldsymbol\lambda}$ will be a fixed point.
Especially, consider the case where ${\sf Gr}$ operates transitively on the components of
${\boldsymbol\lambda}\in\Lambda$, which is equivalent to the condition that for all $\mu=1,\ldots,N$ there exists a
$\pi\in{\mathcal Gr}$
such that $\pi(1)=\mu$. Then it follows that $\hat{\boldsymbol\lambda}={\mathbf 0}$ since this is the only fixed point of the action of
${\sf Gr}$.

This explains why  $\hat{\boldsymbol\lambda}={\mathbf 0}$ in the triangle example 3, where the symmetry group is $D_3$,
isomorphic to ${\mathcal S}_3$. In contrast, in the bow tie example 4, the symmetry group ${\mathcal Gr}$ is generated by the permutations
$(2,3),\; (4,5),$ and $(2,4)(3,5)$. It does not operate transitively on $\{1,\ldots,N\}$ and hence $\hat{\boldsymbol\lambda}$
cannot be determined by pure symmetry considerations. We can only conclude that
$\hat{\lambda}_2=\hat{\lambda}_3=\hat{\lambda}_4=\hat{\lambda}_5$
and thus restrict the domain of possible $\hat{\boldsymbol\lambda}$ to a one-parameter family.

The linear representation $\Pi$ of a permutation $\pi\in{\mathcal S}_N$ also operates on states ${\mathbf s}$ in a natural way
by permuting the spin vectors ${\mathbf s}_\mu,\;\mu=1,\ldots,N$. Let us write this action by ${\mathbf s}\mapsto \Pi\,{\mathbf s}$.
The corresponding action on Gram matrices is given by $G={\mathbf s}\,{\mathbf s}^\top\mapsto \Pi{\mathbf s}\,{\mathbf s}^\top\Pi^\top=
\Pi\,G\,\Pi^\top$.
If $\Pi$ is a symmetry of the spin system, i.~e., commutes with ${\mathbbm J}$,
it follows that the set of ground states of ${\mathbbm J}$ is invariant under the action of $\Pi$. If the ground state is essentially unique,
as in the above triangle example 3, we conclude that for all $\Pi\in{\mathcal Gr}$ there exists an $R\in O(M)$ such that
$\Pi\,{\mathbf s}={\mathbf s}\,R$.
This means that the permutation of the spin numbers can be compensated by some rotation/reflection.
For the triangle example 3 the cyclic shift of the spin numbers
is compensated by a suitable rotation with the angle $2\pi/3$. An equivalent criterion would be that the Gram matrix $G$ commutes with $\Pi$.
In the publication \cite{SL03} ground states with this property have been called ``symmetric ground states".
This means that each ground state has the full symmetry of the whole spin system. In general, this will not be the case:
If ${\mathbf s}$ is a ground state and $\Pi\in{\sf Gr}$ a symmetry, then $\Pi\,{\mathbf s}$ will be another ground state but it need not be
$O(M)$-equivalent to ${\mathbf s}$. If this occurs one says that the symmetry is broken. 

We can prove the existence
of symmetric ground states without any assumption on the symmetry group ${\sf Gr}$:
\begin{theorem}\label{TheoremSym}
 There exists a ground state with Gram matrix $\dot{G}$ such that $\dot{G}=\Pi\,\dot{G}\,\Pi^\top$ for all $\Pi\in{\sf Gr}$.
\end{theorem}
If the spin system is a finite representative of a spin lattice by adopting periodic boundary conditions its symmetry group ${\sf Gr}$
will contain the Abelian subgroup of translations ${\mathcal T}$. ${\mathcal T}$-symmetrical states are sometimes called ``spiral states"
or ``helical states" depending on their dimension.
Hence the above theorem guarantees the existence of $M$-dimensional spiral states that is also investigated in the Ref.~\cite{LT46}--\cite{XW13}
using a completely different method.

We will close this section with an example possessing a large symmetry group and symmetric ground states as well as ground states with broken symmetry.

\vspace{5mm}

\fbox{Example 5: The almost uniform AF octagon ($N=8$)}

\vspace{5mm}

\begin{figure}[ht]
  \centering
    \includegraphics[width=1.0\linewidth]{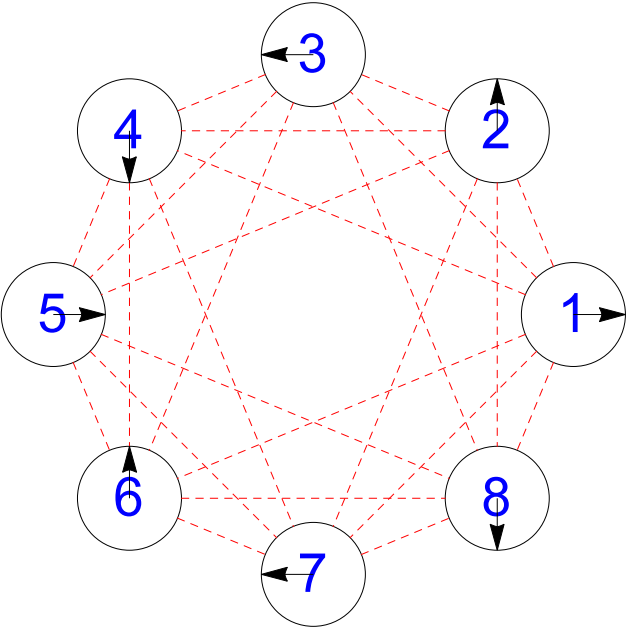}
  \caption[Example 4]
  {The almost uniform AF octagon where each spin is coupled to each other except the opposite one.
  A co-planar, symmetric (``spiral") ground state is indicated by small arrows.}
  \label{FIGEX4}
\end{figure}

The spin system shown in Figure \ref{FIGEX4}  can be described by the undressed ${\mathbbm J}$-matrix
\begin{equation}\label{EX41}
 {\mathbbm J}=\left(
\begin{array}{cccccccc}
 0 & 1 & 1 & 1 & 0 & 1 & 1 & 1 \\
 1 & 0 & 1 & 1 & 1 & 0 & 1 & 1 \\
 1 & 1 & 0 & 1 & 1 & 1 & 0 & 1 \\
 1 & 1 & 1 & 0 & 1 & 1 & 1 & 0 \\
 0 & 1 & 1 & 1 & 0 & 1 & 1 & 1 \\
 1 & 0 & 1 & 1 & 1 & 0 & 1 & 1 \\
 1 & 1 & 0 & 1 & 1 & 1 & 0 & 1 \\
 1 & 1 & 1 & 0 & 1 & 1 & 1 & 0 \\
\end{array}
\right)
\end{equation}

The ground states that can be found numerically by the computer program sketched in the Introduction seem to form a $2$-dimensional
family of $3$-dimensional states having a ground state energy of $E_{min}\approx -16.0\ldots$.
The energy of the collinear state ${\mathbf a}=\uparrow\,\downarrow\,\uparrow\,\downarrow\,\uparrow\,\downarrow\,\uparrow\,\downarrow,$
is exactly $E= -16$, which leads to the conjecture $E_{min}=-16$.

We now apply the present theory to the system under consideration.
The symmetry group of (\ref{EX41}) is $D_8$ and hence operates transitively on the spin sites. According to the above considerations we
conclude $\hat{\boldsymbol\lambda}={\mathbf 0}$. The lowest eigenvalue $j_{min}({\mathbf 0})$ of  ${\mathbbm J}({\mathbf 0})$
is $-2$, corresponding to a ground state energy $E_{min}=-2\times 8 =-16$.
It has a $3$-fold degenerate eigenspace spanned by the $3$ columns of the matrix
\begin{equation}\label{EX4V}
W=\left(
\begin{array}{rrr}
 -1 & -1 & -1 \\
 0 & 0 & 1 \\
 0 & 1 & 0 \\
 1 & 0 & 0 \\
 -1 & -1 & -1 \\
 0 & 0 & 1 \\
 0 & 1 & 0 \\
 1 & 0 & 0 \\
\end{array}
\right).
\end{equation}

\vspace{5mm}
\begin{figure}[ht]
  \centering
    \includegraphics[width=1.0\linewidth]{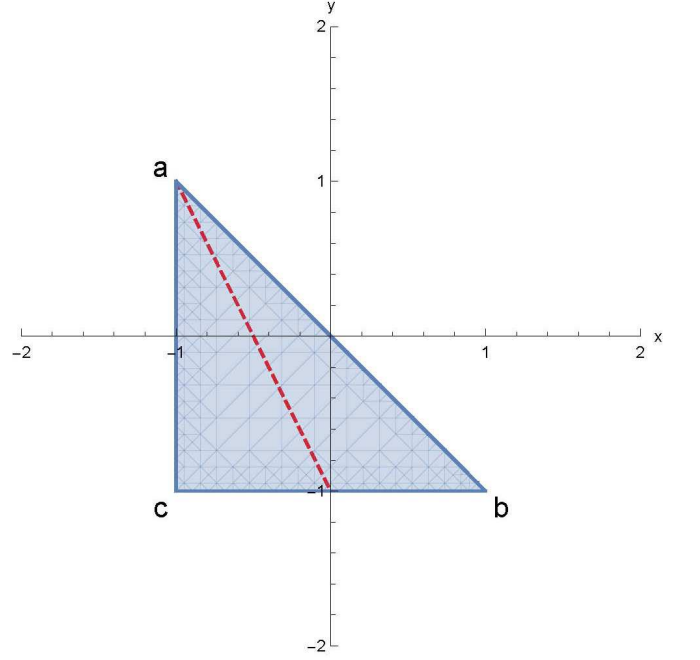}
  \caption[Example 4R]
  {The triangular region of parameters $(x,y)$ leading to a solution $\Delta(x,y)\ge0$ of the ADE and
  hence to ground states of the spin system displayed in Figure \ref{FIGEX4}.
  The dashed red line indicates those point that lead to symmetric ground states. The three vertices of the triangle
  correspond to collinear ground states ${\mathbf a}$, ${\mathbf b}$ and ${\mathbf c}$. }
  \label{FIGEX4R}
\end{figure}

The corresponding ADE (\ref{DSG6}) has the $2$-parameter family of solutions
\begin{equation}\label{EX42}
\Delta(x,y)=
\left(
\begin{array}{ccc}
 1 & x & y \\
 x & 1 & -x-y-1 \\
 y & -x-y-1 & 1 \\
\end{array}
\right)
\;.
\end{equation}
Moreover, $\det \Delta(x,y)= -2 (x+1) (y+1) (x+y)$. From this it follows that the domain $\Delta(x,y)\ge 0$ is formed by the
triangle in the $x-y$-plane bounded by the lines $x=-1,\; y=-1,$ and $x+y=0$, see Figure \ref{FIGEX4R}.
In the interior of the triangle we have $3$-dimensional ground states
parametrized by $x$ and $y$ such that the corresponding Gram matrix has the form\\

\onecolumngrid

\begin{equation}\label{EX43}
 G(x,y)=\left(
\begin{array}{cccccccc}
 1 & x & y & -x-y-1 & 1 & x & y & -x-y-1 \\
 x & 1 & -x-y-1 & y & x & 1 & -x-y-1 & y \\
 y & -x-y-1 & 1 & x & y & -x-y-1 & 1 & x \\
 -x-y-1 & y & x & 1 & -x-y-1 & y & x & 1 \\
 1 & x & y & -x-y-1 & 1 & x & y & -x-y-1 \\
 x & 1 & -x-y-1 & y & x & 1 & -x-y-1 & y \\
 y & -x-y-1 & 1 & x & y & -x-y-1 & 1 & x \\
 -x-y-1 & y & x & 1 & -x-y-1 & y & x & 1 \\
\end{array}
\right).
\end{equation}

\twocolumngrid
At the three edges of the triangle, i.~e.~, for the values $x=-1,\;-1<y<1$ or $y=-1,\;-1<x<1$ or $-1<x=-y<1$ we have co-planar ground states
as, for example, indicated in the Figure \ref{FIGEX4}. The three vertices of the triangle correspond to the collinear ground states \\
${\mathbf a}=\uparrow\,\downarrow\,\uparrow\,\downarrow\,\uparrow\,\downarrow\,\uparrow\,\downarrow,$ ,
${\mathbf b}=\uparrow\,\uparrow\,\downarrow\,\downarrow\,\uparrow\,\uparrow\,\downarrow\,\downarrow,$
and ${\mathbf c}=\uparrow\,\downarrow\,\downarrow\,\uparrow\,\uparrow\,\downarrow\,\downarrow,\,\uparrow$.

In general $G(x,y)$ does not commute with the cyclic shift matrix $C$ that is the linear representation of the cyclic permutation
$\pi=(12345678)\in{\mathcal S}_8$.
 This can be confirmed by inspection of (\ref{EX43}),
since $[G(x,y),C]=0$ means that the secondary diagonals of $G(x,y)$ should be constant, even if they are extended periodically.
Matrices with this property are called ``circulants", see \cite{LT85}.
A detailed calculation yields $C^{-1}\,  G(x,y)\,C= G(-1-x-y,y)$. This implies that only the points $(x,y)$ of the triangle
satisfying the equation $y=-1-2x$ lead to circulant Gram matrices $G(x,y)$ and hence to symmetric ground states.
The special symmetric co-planar ground state corresponding to the point $(x=0,y=-1)$ is indicated in Figure \ref{FIGEX4}
by small arrows attached to the spin sites. Only the collinear ground state ${\mathbf a}$ is symmetric, whereas
${\mathbf b}$ and ${\mathbf c}$ are interchanged by the cyclic shift.

The present example serves to illustrate the following points:
\begin{itemize}
  \item It is an example of a system with a large symmetry group and ground states with broken symmetry as well as symmetric ground states,
  \item It has ground states of all physical dimensions $1,2$ and $3$,
  \item It is frustrated and has nevertheless collinear ground states,
  \item It has an additional degeneracy of degree $2$ that is not due to an independent rotation of the spin vectors of some subgroup,
  \item It shows how the present theory works for an example of medium complexity and how it extends the information
  available by numerical calculations.
\end{itemize}

\section{Example of a non-standard system}\label{sec:NS}

Recall that the minimal eigenvalue $j_{min}({\boldsymbol\lambda})$ of ${\mathbbm J}({\boldsymbol\lambda})$
assumes its maximum $\hat{\jmath}$ at a unique point $\hat{\boldsymbol\lambda}\in \Lambda$ and that there exists
at least one ground state ${\mathbf s}$ that is living on the eigenspace of $({\mathbbm J}({\boldsymbol\lambda}),\hat{\jmath})$.
The spin system has been called ``standard" iff at least one ground state of this kind has a dimension $\mbox{dim}({\mathbf s})\le 3$.
In this section we will provide an example of a spin system with $N=10$ that has an essentially unique ground state
of dimension $4$ and a ground state energy $E_{min}=N\, \hat{\jmath}$.
Hence the physical ground states with dimension at most $3$ will have a larger energy.

The ${\mathbbm J}$-matrix of the example is too complicated to be displayed here. We will rather describe the procedure how to obtain it.
In some sense we have to invert the process of finding ground states if the spin system is given:
We start with a suitable ground state and construct a spin system that possesses this very ground state.
In view of Proposition \ref{PropADE} the intended unique ground state
${\mathbf s}$ should have $N$ row vectors ${\mathbf s}_\mu$ of length $M=4$
such that the corresponding projectors $P_\mu$ span ${\mathcal S}{\mathcal M}(4)$. Hence we need at least $N=M(M+1)/2=10$ such row vectors.
The following choice satisfies these requirements:
\begin{equation}\label{NS1}
 {\mathbf s}=\left(
\begin{array}{cccc}
 \frac{1}{\sqrt{2}} & \frac{1}{\sqrt{2}} & 0 & 0 \\
 \frac{1}{\sqrt{2}} & 0 & \frac{1}{\sqrt{2}} & 0 \\
 \frac{1}{\sqrt{2}} & 0 & 0 & \frac{1}{\sqrt{2}} \\
 0 & \frac{1}{\sqrt{2}} & \frac{1}{\sqrt{2}} & 0 \\
 0 & \frac{1}{\sqrt{2}} & 0 & \frac{1}{\sqrt{2}} \\
 0 & 0 & \frac{1}{\sqrt{2}} & \frac{1}{\sqrt{2}} \\
 1 & 0 & 0 & 0 \\
 0 & 1 & 0 & 0 \\
 0 & 0 & 1 & 0 \\
 0 & 0 & 0 & 1 \\
\end{array}
\right).
\end{equation}
The $4$ columns of ${\mathbf s}$ span a $4$-dimensional subspace of ${\mathbbm R}^{10}$.
We calculate the projector ${\mathbbm Q}$  onto this subspace and set
\begin{equation}\label{NS2}
 {\mathbbm J}({\boldsymbol\lambda})\equiv -6 {\mathbbm Q} +4 ({\mathbbm 1}-{\mathbbm Q})
 \;.
\end{equation}
One easily checks that ${\mathbbm J}({\boldsymbol\lambda})$ is a symmetric matrix and
$\mbox{Tr }(-6 {\mathbbm Q} +4 ({\mathbbm 1}-{\mathbbm Q}))=0$, hence this matrix can indeed be written
as ${\mathbbm J}({\boldsymbol\lambda}),\,{\boldsymbol\lambda}\in\Lambda$.
The lowest eigenvalue $-6$ of ${\mathbbm J}({\boldsymbol\lambda})$ will be $4$-times degenerate
with the projector ${\mathbbm Q}$ onto the corresponding eigenspace $S$.
The $10\times 4$-matrix $W$ the columns of which span $S$ can be chosen as $W={\mathbf s}$.
By construction, the ADE  has the unique solution $\Delta={\mathbbm 1}$
and (\ref{NS1})
is the unique $4$-dimensional ground state up to rotational/reflectional degeneracy and will have a ground state energy
$E_{min}= 10\times (-6) =-60$. We have numerically determined  $3$-dimensional spin configurations with the
lowest energy $E_0^{(3)}$ by the method sketched in the Introduction.
The result was $E_0^{(3)}=-59.17279762005$, where all decimals are obtained in a reproducible manner.
Hence $E_0^{(3)}$ lies only slightly but definitively above $E_{min}=-60$.
Thus the claim that the corresponding spin system is not a standard one is also numerically confirmed.

However, the mere effort to find such an example may be considered as an argument to expect
that in practice most spin systems will be standard ones and hence the concentration on standard
systems in this paper seems to be justified.

The above method can be extended to yield spin systems with $N=M(M+1)/2$ spins  that possess essentially unique
ground states with dimension $M$ for any $M=2,3,\ldots$.
The case $M=3,\,N=6$ is particularly interesting even if the resulting spin system
is standard. The corresponding dressed ${\mathbbm J}$-matrix has the form:
\begin{equation}\label{NS3}
{\mathbbm J}({\boldsymbol\lambda})=
\left(
\begin{array}{cccccc}
 1 & -2 & -2 & -4 \sqrt{2} & -4 \sqrt{2} & 2 \sqrt{2} \\
 -2 & 1 & -2 & -4 \sqrt{2} & 2 \sqrt{2} & -4 \sqrt{2} \\
 -2 & -2 & 1 & 2 \sqrt{2} & -4 \sqrt{2} & -4 \sqrt{2} \\
 -4 \sqrt{2} & -4 \sqrt{2} & 2 \sqrt{2} & -1 & 2 & 2 \\
 -4 \sqrt{2} & 2 \sqrt{2} & -4 \sqrt{2} & 2 & -1 & 2 \\
 2 \sqrt{2} & -4 \sqrt{2} & -4 \sqrt{2} & 2 & 2 & -1 \\
\end{array}
\right).
\end{equation}

\section{Proofs}\label{sec:P}

In order to state rigorous proofs for the various propositions and theorems of the main sections we first have to
explicitly formulate some general, rather trivial assumptions about the spin systems under consideration.
\begin{ass}\label{Ass1}
 The number of spins $N$ satisfies $N\ge 3$.
\end{ass}
This assumption is sensible since the case $N=2$ is completely treated in Example 1.
\begin{ass}\label{Ass2}
 The real, symmetric $N\times N$ matrix ${\mathbbm J}$ has some non-vanishing non-diagonal elements.
\end{ass}
The second assumption implies that for arbitrary gauges ${\boldsymbol\lambda}\in\Lambda$,  ${\mathbbm J}({\boldsymbol\lambda})$
is never the zero matrix and hence has some eigenvalue
$j_\alpha({\boldsymbol\lambda})\neq 0$. Since the trace of ${\mathbbm J}({\boldsymbol\lambda})$ vanishes, the minimal eigenvalue must be
negative, $j_{min}({\boldsymbol\lambda})<0$, and hence the following holds:
\begin{lemma}\label{Lemma0}
\begin{equation}\label{AP0}
\hat{\jmath}=\sup \{j_{min}({\boldsymbol\lambda})|{\boldsymbol\lambda}\in\Lambda\}\le 0\;.
\end{equation}
\end{lemma}
Actually, $\hat{\jmath}<0$ since the supremum is assumed due to Proposition \ref{Prop2}, but for the proof of
Lemma \ref{Lemma4} we will only use Lemma \ref{Lemma0}.

Except some finite sets as the symmetric group ${\mathcal S}_N$ etc.~all sets considered in this paper
are subsets of some ${\mathbbm R}^n$ or spaces homeomorphic to ${\mathbbm R}^n$.
All topological concepts used for these sets
hence refer to the standard topology of ${\mathbbm R}^n$ or the corresponding topology inherited by its subsets.

A state ${\mathbf s}$ of a spin system has been defined in the main text as an $N\times M$-matrix such that its
rows ${\mathbf s}_\mu,\;\mu=1,\ldots,N$ are unit vectors of ${\mathbbm R}^M$. For the present section we slightly
modify this definition by considering equivalence classes of such matrices.
Two $N\times M_i$-matrices ${\mathbf s}^{(i)},\;i=1,2,$ are considered as equivalent iff, in the case $M_1\le M_2$,
the matrix ${\mathbf s}^{(2)}$ is obtained from ${\mathbf s}^{(1)}$ by padding $M_2-M_1$ zero columns, and analogously
in the case $M_2\le M_1$. Of course, for each such equivalence class there exists a natural representative, namely
the matrix with a minimal $M$, possessing no zero columns at the right hand side. If we represent a state simply by a matrix
in what follows we tacitly use this natural representative.
Note also that the above equivalence relation induces a natural embedding ${\mathcal P}_M\subset {\mathcal P}_{M'}$
if $M<M'$.

\subsection{The Lagrange variety approach}\label{sec:PLV}

\noindent{\bf Proof of Proposition \ref{PropV}}\\
Define
$A\equiv {\mathbbm J}({\boldsymbol\lambda})-x\,{\mathbbm 1}$ such that
$p({\boldsymbol\lambda },x)=\det\, A$.
According to the assumption $\frac{\partial p({\boldsymbol\lambda },x)}{\partial x}=0$
the eigenvalue $0$ of $A$ is at least twofold degenerate.
Let $A^{(1)}$ be the matrix resulting from deleting the first row and the first column of $A$, analogously for $A^{(N)}$.
According to Cauchy's interlacing theorem $A^{(1)}$ and $A^{(N)}$ have also the eigenvalue $0$ and hence
\begin{equation}\label{APV1}
 \det\,A^{(1)}=\det\,A^{(N)}=0
 \;.
\end{equation}
We regard $\det A$ as a polynomial in the variables $A_{ij}$ and write
\begin{equation}\label{APV2}
 \det A=A_{11}R_1+A_{NN}R_N+A_{11}A_{NN}R_{1N}+R_0
 \;,
\end{equation}
such that the factors $R_0,\,R_1,\,R_N,\,R_{1N}$ do not contain $A_{11}$ or $A_{NN}$.
The Laplacian determinant expansion by minors yields
\begin{eqnarray}\label{APV3a}
 \det A^{(1)}&=&R_1+A_{NN}R_{1N},\\
 \label{APV3b}
 \det A^{(N)}&=&R_N+A_{11}R_{1N}
 \;.
\end{eqnarray}

We want to show that $\frac{\partial p({\boldsymbol\lambda},x)}{\partial\lambda_1}=0$. Obviously,
\begin{equation}\label{APV8}
 A_{11}=\lambda_1-x,\;A_{NN}=-\sum_{i=1}^{N-1}\lambda_i\;- x
 \;,
\end{equation}
are the only matrix elements of $A$ containing $\lambda_1$. Hence the corresponding partial derivative of (\ref{APV2}) yields
\begin{eqnarray}\nonumber
  \frac{\partial p({\boldsymbol\lambda},x)}{\partial\lambda_1} &=& \frac{\partial\det A}{\partial\lambda_1} \\
  \nonumber
   &=& R_1-R_N+\frac{\partial}{\partial\lambda_1}\left( A_{11}A_{NN}\right)R_{1N}\\
  \nonumber
   &=& \left( R_1+A_{NN}R_{1N}\right)-\left( R_N+A_{11}R_{1N}\right)\\
   \label{APV9}
   &=&\det A^{(1)}-\det A^{(N)}=0
   \;,
   \end{eqnarray}
by (\ref{APV3a}), (\ref{APV3b}) and (\ref{APV1}). The proof of $\frac{\partial p({\boldsymbol\lambda},x)}{\partial\lambda_i}=0,\;$
 $i=2,\ldots,N-1$
is analogous. \hfill$\Box$\\

\noindent{\bf Proof of Proposition \ref{Prop1}}\\
The claim is equivalent to the sub-graph $J_\le\equiv\{({\boldsymbol\lambda},x)\in\Lambda\times{\mathbb R}\,|\,x\le j_{min}({\boldsymbol\lambda})\}$
being a convex set. The latter holds since $J_\le$ is the intersection of the family of convex closed half-spaces
$ H_\varphi\equiv\{({\boldsymbol\lambda},x)\in\Lambda\times{\mathbb R}\,|\,x\le \langle\varphi|{\mathbbm J}({\boldsymbol\lambda})\varphi\rangle\} $
where $\varphi\in{\mathbb R}^N $ and $||\varphi||=1$. \hfill$\Box$\\

\noindent{\bf Proof of Proposition \ref{Prop2}}\\
It follows immediately from  Proposition \ref{Prop1} that $\widehat{J}$ will be closed and convex since it is the intersection of
two closed convex sets,
$\widehat{J}=J_\le \,\cap\,
\{({\boldsymbol\lambda},x)\in\Lambda\times{\mathbbm R}|x=\hat{\jmath}\}$.
The harder part is to prove that $\widehat{J}$ is non-empty.
To this end we will state some auxiliary lemmas. We will identify $\Lambda$ with ${\mathbb R}^{N-1}$ via projection onto the first $N-1$
components of ${\boldsymbol\lambda}\in\Lambda$, recalling that
\begin{equation}\label{AP1}
\lambda_N = -\sum_{\nu=1}^{N-1}\lambda_\nu
\;,
\end{equation}
compare (\ref{D4}). W.~r.~t.~this identification we will use the notation $\|{\boldsymbol\lambda}\|$ for the norm of ${\boldsymbol\lambda}\in\Lambda$.
According to a general theorem, a real, continuous function defined on a compact set assumes its supremum.
In our case, $j_{min}$ is continuous, but it is defined on the subspace
$\Lambda$ that is not compact. Hence we want to restrict $j_{min}$ to a compact subset of the form
$\{{\boldsymbol\lambda}\in\Lambda|\,||{\boldsymbol\lambda}||\le C\}$ in such a way that its supremum remains unchanged.
To this end we choose some real number $C$ satisfying
\begin{equation}\label{AP2}
C\ge N^2\,\left|\hat{\jmath}\right|+N
\;,
\end{equation}
and state the following
\begin{lemma}\label{Lemma3}
If $\|{\boldsymbol\lambda}\|>C$ then $j_{min}({\boldsymbol\lambda})<-\frac{C}{N^2}$.
\end{lemma}
{\bf Proof of Lemma \ref{Lemma3}}\\
By the Rayleigh-Ritz variational principle,
\begin{equation}\label{AP3}
j_{min}({\boldsymbol\lambda})\le \lambda_\mu \mbox{   for all  }\mu=1,\ldots,N
\;.
\end{equation}
If for all $\nu=1,\ldots,N-1$ we would have $\left| \lambda_\nu\right|\le \frac{C}{N}$, the triangle inequality would imply
\begin{equation}\label{AP3a}
\| {\boldsymbol\lambda} \| \le \sum_{\nu=1}^{N-1} \left| \lambda_\nu \right| \le (N-1)\frac {C}{N}< C \;,
\end{equation}
which contradicts the assumption $\| {\boldsymbol\lambda}\|>C$. Hence at least one $\lambda_\nu,\;\nu=1,\ldots,N-1$ satisfies
\begin{equation}\label{AP4}
\left|\lambda_\nu \right|>\frac{C}{N}
\;.
\end{equation}
We will fix this $\nu$ for the rest of the proof. If $\lambda_\nu<0$ the claim follows by
since $j_{min}({\boldsymbol\lambda})\stackrel{(\ref{AP3})}{\le} \lambda_\nu \stackrel{(\ref{AP4})}{<}- \frac{C}{N}<-\frac{C}{N^2}$,
using $N\ge 3$ in the last step.
Hence we may assume
\begin{equation}\label{AP4a}
\lambda_\nu>0
\;,
\end{equation}
since $\lambda_\nu=0$ is excluded by (\ref{AP4}).
Moreover, (\ref{AP4}) implies
\begin{equation}\label{AP4a}
\lambda_\nu >\frac{C}{N}
\;.
\end{equation}

Let us define the set
\begin{equation}\label{AP5}
{\mathcal K}\equiv \{\kappa=1,\ldots,N-1|\lambda_\kappa<0\}
\;,
\end{equation}
which may be empty, and
\begin{equation}\label{AP5a}
K\equiv\left|{\mathcal K}\right|
\;.
\end{equation}
Since (\ref{AP4a}) implies  $\nu\notin {\mathcal K}$, we have
\begin{equation}\label{AP6}
0\le K\le N-2
\;.
\end{equation}
If for some $\kappa\in{\mathcal K}$ we would have $\lambda_\kappa < - \frac{C}{N^2}$
then the claim would follow by (\ref{AP3}).
Hence we may assume $\lambda_\kappa\ge -\frac{C}{N^2}$ and consequently
\begin{equation}\label{AP7}
\left|\lambda_\kappa\right|\le \frac{C}{N^2} \mbox{   for all    }\kappa\in{\mathcal K}
\;.
\end{equation}
It follows that
\begin{eqnarray}\label{AP8a}
\lambda_N&\stackrel{(\ref{AP1})}{=}&-\sum_{\mu=1}^{N-1}\lambda_\mu\\
\label{AP8b}
&\stackrel{(\ref{AP5})}{\le}&-\lambda_\nu-\sum_{\kappa\in{\mathcal K}}\lambda_\kappa\\
\label{AP8c}
&\stackrel{(\ref{AP5})}{=}&-\lambda_\nu+\sum_{\kappa\in{\mathcal K}}\left|\lambda_\kappa\right|\\
\label{AP8d}
&\stackrel{(\ref{AP4a})(\ref{AP7})}{<}&-\frac{C}{N}+K\,\frac{C}{N^2}=\frac{C}{N^2}(K-N)\\
\label{AP8e}
&\stackrel{(\ref{AP6})}{\le}&-\frac{2C}{N^2}<-\frac{C}{N^2}
\;,
\end{eqnarray}
from which the claim follows by (\ref{AP3}).
This completes the proof of Lemma \ref{Lemma3}.\hfill$\Box$\\
Next we prove that the restriction of $j_{min}$ to the compact ball with radius $C$ does not change its
supremum:\\

\begin{lemma}\label{Lemma4}
$\hat{\jmath}=\sup \{j_{min}({\boldsymbol\lambda})|{\boldsymbol\lambda}\in\Lambda\mbox{ and } ||{\boldsymbol\lambda}||\le C\}\;.$
\end{lemma}

\noindent{\bf Proof of Lemma \ref{Lemma4}}\\
Recall that $\hat{\jmath}$ was defined as
$\hat{\jmath}=\sup \{j_{min}({\boldsymbol\lambda})|{\boldsymbol\lambda}\in\Lambda\}\;.$
This means that for all $\varepsilon>0$ there exists a ${\boldsymbol\lambda}\in\Lambda$ such that
\begin{equation}\label{AP7a}
j_{min}({\boldsymbol\lambda})>\hat{\jmath}-\varepsilon
\;.
\end{equation}
 We may additionally require
\begin{equation}\label{AP8}
\varepsilon < \frac{1}{N}
\;.
\end{equation}
In order to derive a contradiction assume $||{\boldsymbol\lambda}||>C$. Then, by virtue of Lemma \ref{Lemma3},
\begin{eqnarray}\label{AP9a}
j_{min}({\boldsymbol\lambda})&<&-\frac{C}{N^2}\stackrel{(\ref{AP2})}{\le} -|\hat{\jmath}|-\frac{1}{N}\\
\label{AP9b}
&\stackrel{(\ref{AP0})}{=}&\hat{\jmath}-\frac{1}{N}\stackrel{(\ref{AP8})}{<}\hat{\jmath}-\varepsilon
\;,
\end{eqnarray}
which contradicts (\ref{AP7a}). Hence $||{\boldsymbol\lambda}||\le C$ holds. This proves:
For all $0<\varepsilon<\frac{1}{N}$ there exists a ${\boldsymbol\lambda}\in\Lambda$ such that $||{\boldsymbol\lambda}||\le C$
and $j_{min}({\boldsymbol\lambda})>\hat{\jmath}-\varepsilon$.
Hence the claim of Lemma \ref{Lemma4} is satisfied.\hfill$\Box$\\

Now Proposition \ref{Prop2} follows since $\{{\boldsymbol\lambda}\in\Lambda|\,||{\boldsymbol\lambda}||\le C\}$
is compact and $j_{min}$ is continuous, see corollary \ref{Cor1}.

The proof of Lemma \ref{Lemma4} also shows that $\widehat{J}$ is closed and bounded, hence compact.\hfill$\Box$\\

\subsection{Degeneracy}\label{sec:PD}

\noindent{\bf Proof of Proposition \ref{Prop5}}\\
We consider the system of vectors $({\mathbf s}_\mu)_{\mu=1,\ldots,N}$ where ${\mathbf s}_\mu\in{\mathbb R}^M$
and perform the process of Gram-Schmidt orthogonalization
resulting in the orthonormal basis $({\mathbf r}_\nu)_{\nu=1,\ldots,M}$ of ${\mathbb R}^M$. Since $M\le N$
some ${\mathbf s}_\mu$ may be linear combinations of the ${\mathbf s}_\alpha,\; \alpha=1,\ldots,\mu-1$ and will not contribute
to the orthonormal basis $({\mathbf r}_\nu)_{\nu=1,\ldots,M}$. Let $({\mathbf s}_\mu)_{\mu\in{\mathcal A}}$ be this family
of not contributing vectors and $({\mathbf s}_\mu)_{\mu\in{\mathcal B}}$  its complement, such that $|{\mathcal B}|=M$ and
$|{\mathcal A}|=N-M$. We rearrange both sets of row vectors into matrices ${\mathbf a}, {\mathbf b}$ such that
${\mathbf b}$ has $M$ rows and ${\mathbf a}$ has $N-M$ rows.
The vectors ${\mathbf b}_\mu,\,\mu=1,\ldots,M$ are unique linear combinations of the orthonormal basis vectors
${\mathbf r}_\nu,\,\nu=1,\ldots,M$:
\begin{equation}\label{AP11}
{\mathbf b}_\mu=\sum_{\nu=1}^M\,\sigma_{\mu\nu}\,{\mathbf r}_\nu,
\end{equation}
or, in matrix notation
\begin{equation}\label{AP12a}
{\mathbf b}={\boldsymbol\sigma}\,{\mathbf r}
\;,
\end{equation}
and, equivalently,
\begin{equation}\label{AP12b}
{\mathbf b}^\top={\mathbf r}^\top\,{\boldsymbol\sigma}^\top
\;.
\end{equation}
Note that the $\sigma_{\mu\nu}$ can be solely expressed in terms of scalar products ${\mathbf s}_\mu\cdot{\mathbf s}_\nu=G_{\mu\nu}$
and thus are the same for all spin configurations with the same Gram matrix.
Further, ${\mathbf r }\in O(M)$ since the rows of ${\mathbf r}$ form an orthonormal basis of ${\mathbb R}^M$.\\
We now consider two spin configurations ${\mathbf s}^{(i)},\;i=1,2$ with the same Gram matrix
\begin{equation}\label{AP16}
G={\mathbf s}^{(1)}\,{\mathbf s}^{(1)\top} = {\mathbf s}^{(2)}\,{\mathbf s}^{(2)\top}
\;,
\end{equation}
and rewrite (\ref{AP12b}) in the form
\begin{equation}\label{AP17}
{\mathbf b}^{(1)\top} ={\mathbf  r}^{(1)\top} \,{\boldsymbol\sigma}^\top,\quad
{\mathbf b}^{(2)\top} ={\mathbf  r}^{(2)\top} \,{\boldsymbol\sigma}^\top
\;,
\end{equation}
where we have used the fact that the matrix ${\boldsymbol\sigma}$
is the same for both spin configurations, see above. We conclude that
\begin{eqnarray}\label{AP18a}
 {\mathbf b}^{(2)\top} &=& {\mathbf  r}^{(2)\top} \,{\boldsymbol \sigma}^\top ={\mathbf  r}^{(2)\top} \,
  {\mathbf  r}^{(1)}\,{\mathbf b}^{(1)\top}\\
  \label{AP18b}
   &\equiv& R\;{\mathbf b}^{(1)\top}
   \;,
\end{eqnarray}
with some rotation/reflection $R\in O(M)$. This proves the claim for the vectors
${\mathbf b}_\mu^{(1)},\,{\mathbf b}_\mu^{(2)},\; \mu=1,\ldots,M$.
For the remaining vectors ${\mathbf a}^{(i)}_\mu,\;\mu=1,\ldots,N-M,\,i=1,2$ the statement
analogous to (\ref{AP18b}) follows from the representation
\begin{equation}\label{AP19}
{\mathbf a}_\mu^{(i)}=\sum_{\nu=1}^M\, \tau_{\mu\nu}\,{\mathbf b}_\nu^{(i)},\;\mu=1,\ldots,N-M
 \;,
\end{equation}
and the fact that the coefficients $\tau_{\mu\nu}$ can be expressed solely in terms of scalar products
${\mathbf s}_\mu^{(i)}\cdot{\mathbf s}_\nu^{(i)}$ and hence do not depend on $i$. \hfill$\Box$\\

\noindent{\bf Proof of Proposition \ref{PropC}}\\
Let $S\subset S'$ be the subspace generated by the set
\begin{equation}\label{APC1}
\Sigma\equiv \{{\mathbf s}_i \left|  {\mathbf s}\in{\mathcal P}_{M',S'},\,i=1,\ldots,M'\right.\}
\;.
\end{equation}
$S$ is already generated by a finite subset $\Sigma_{fin}\subset\Sigma$ that can be chosen to be of the form
\begin{equation}\label{APC2}
\Sigma_{fin}= \{{\mathbf s}_i^{(j)} \left|  {\mathbf s}^{(j)}\in{\mathcal P}_{M',S'},\,
j=1,\ldots,m,\,i=1,\ldots,M'\right.\}
\;,
\end{equation}
where $m\ge 1$ is some integer. In other words, $S$ is generated by the columns of a finite set of spin configurations.
Let $\bar{\mathbf s}$ be the $N\times \bar{M}$-matrix resulting from the horizontal
juxtaposition of the matrices $\frac{1}{\sqrt{m}}\,{\mathbf s}^{(j)},\;j=1,\ldots,m$, hence $\bar{M}=m\,M'$.
It follows that for $\mu=1,\ldots,N$
\begin{equation}\label{APC3}
\bar{\mathbf s}_\mu \cdot\bar{\mathbf s}_\mu =\sum_{i=1}^{\bar{M}}{\mathbf s}_{\mu i}^2
=\sum_{j=1}^{m}\sum_{i=1}^{M'}\left(\frac{1}{\sqrt{m}}\,{\mathbf s}_{\mu i}^{(j)} \right)^2
=\sum_{j=1}^{m}\frac{1}{m}=1.
\end{equation}
Hence $\bar{\mathbf s}\in{\mathcal P}_{\bar{M},S}$. Since the $\bar{M}$ columns of $\bar{\mathbf s}$
span $S$ its dimension (rank) is $\mbox{dim }\bar{\mathbf s}=\mbox{dim }S=M$.

Now consider the subspace
$T\subset{\mathbbm R}^{\bar{M}}$
spanned by the $N$ rows $\bar{\mathbf s}_\mu$ of $\bar{\mathbf s}$.
Its dimension is $\mbox{dim }T=\mbox{rank }\bar{\mathbf s}=M$. Further, let $T'$ be the subspace of ${\mathbbm R}^{\bar{M}}$
spanned by the first $M$ elements ${\mathbf e}_\mu,\,\mu=1,\ldots,M$ of the standard basis of  ${\mathbbm R}^{\bar{M}}$.
Since $T$ and $T'$ have the same dimension there exists an $R\in O(\bar{M})$ that maps $T'$ onto $T$.
Hence $R^{-1}=R^\top$ maps $T$ onto $T'$ and we have the following implications:
\begin{eqnarray}\label{APC4a}
  R^\top \bar{\mathbf s}_\mu^\top & \in & T' \mbox{ for all }\mu =1,\ldots N,\\
  \label{APC4b}
 \left(\bar{\mathbf s}_\mu\,R \right)^\top & \in & T' \mbox{ for all }\mu =1,\ldots N, \\
\label{APC4c}
 \mbox{ all rows of }&\bar{{\mathbf s}}\,R&
 \mbox{ lie in } T'  \;,\\
  \nonumber
 \left(\bar{\mathbf s}\,R \right)_{\mu i} &=& 0 \mbox{ for all } i=M+1,\ldots,\bar{M}\\
 \label{APC4d}
 && \mbox{ and } \mu=1,\ldots, N  \;.
\end{eqnarray}
It follows that ${\mathbf s}\equiv \bar{\mathbf s}\,R \in {\mathcal P}_{M,S}$ with respect to the natural embedding
${\mathcal P}_{M,S}\subset {\mathcal P}_{\bar{M},S}$ following from the remarks at the outset of this section.  \hfill$\Box$\\

\noindent{\bf Proof of Proposition \ref{ProppM}}\\
We assume that $\mbox{dim }({\mathbf s})=M $ and, without loss of generality, that the first $M$
spin vectors ${\mathbf s}_\mu,\;\mu=1,\ldots,M$ already span ${\mathbbm R}^M$ and, moreover, that
${\mathbf s}_1=(1,0,\ldots,0)^\top,\;{\mathbf s}_2=(\ast,{\boldsymbol\ast},\ldots,0)^\top,\;\ldots,
{\mathbf s}_M=(\ast,\ast,\ldots,{\boldsymbol\ast})^\top$.
Here $\ast$ denotes some real number and ${\boldsymbol\ast}$ some non-vanishing real number.
The latter can be achieved by choosing a suitable
rotation/reflection $R\in O(M)$ and replacing $R{\mathbf s}_\mu$ by ${\mathbf s}_\mu$. It follows that
the corresponding projections $P_\mu,\;\mu=1,\ldots,M$ are $\mu\times\mu$-matrices with non-vanishing entries
$\left(P_\mu\right)_{\mu\mu}$ and padded with zeroes to obtain an $M\times M$-matrix.
Hence the set of $P_\mu,\;\mu=1,\ldots,M$ is linearly independent and $p\ge M$.
The total number of projections $P_\mu,\;\mu=1,\ldots,N$ is $N$ and hence $p\le N$.      \hfill$\Box$\\

\subsection{Fusion}\label{sec:PF}
\noindent{\bf Proof of Proposition \ref{PropF1}}\\
(i)\quad Due to the construction of the fusion of two states it follows that $H_i({\mathbf s}^{(i)})=H({\mathbf S}^{(i)})$ for $i=1,2$,
and $H({\mathbf S}^{(2)})=H(\bar{\mathbf s}^{(2)})$ by the $O(M_1+M_2)$-invariance of $H$.
Hence $H({\mathbf s})=H_1({\mathbf s}^{(1)})+H_2({\mathbf s}^{(2)})$. Now assume that ${\mathbf s}$ is not a ground state of $H$, i.~e.~,
that there exists a state $\tilde{\mathbf s}$ with $H(\tilde{\mathbf s})<H({\mathbf s})$. Let
$\tilde{\mathbf s}^{(1)}_\mu\equiv \tilde{\mathbf s}_\mu$ for $\mu=1,\ldots ,N_1$ and
$\tilde{\mathbf s}^{(2)}_\mu\equiv \tilde{\mathbf s}_\mu$ for $\mu=N_1,\ldots ,N$
such that $H(\tilde{\mathbf s})=H_1(\tilde{\mathbf s}^{(1)})+H_2(\tilde{\mathbf s}^{(2)})$.
It follows that either
$H_1(\tilde{\mathbf s}^{(1)})<H_1({\mathbf s}^{(1)})$ or $H_2(\tilde{\mathbf s}^{(2})<H_2({\mathbf s}^{(2)})$
which contradicts the assumption that the states ${\mathbf s}^{(i)}$ are ground states for $i=1,2$.\\
(ii) \quad Let ${\mathbf s}$ be an $N\times M$-matrix that is a ground state of $H$ and set, similarly as in (i),
${\mathbf s}^{(1)}_\mu\equiv {\mathbf s}_\mu$ for $\mu=1,\ldots ,N_1$ and
${\mathbf s}^{(2)}_\mu\equiv {\mathbf s}_\mu$ for $\mu=N_1,\ldots ,N$
such that $H({\mathbf s})=H_1({\mathbf s}^{(1)})+H_2({\mathbf s}^{(2)})$.
In order to derive a contradiction assume that ${\mathbf s}^{(1)}$ is not a
ground state of $H_1$, i.~e.~, that there exists an $\tilde{\mathbf s}^{(1)}$ such that
$H_1(\tilde{\mathbf s}^{(1)})<H_1({\mathbf s}^{(1)})$. Choose $R\in O(M)$ such that
$R\, \tilde{\mathbf s}^{(1)}_{N_1} = {\mathbf s}^{(2)}_{N_1}$ and define
$\bar{\mathbf s}^{(1)}_\mu \equiv R\, \tilde{\mathbf s}^{(1)}_{\mu}$ for $\mu=1,\ldots,N_1$. By the $O(M)$-invariance
of $H_1$ we have $H_1(\bar{\mathbf s}^{(1)})=H_1(\tilde{\mathbf s}^{(1)})$. Then the definition
\begin{equation}\label{PF1}
\bar{\mathbf s}_\mu=  \left\{\begin{array}{l@{\quad:\quad}l}
 \bar{\mathbf s}^{(1)}_\mu& 1\le \mu\le N_1,\\
 {\mathbf s}^{(2)}_\mu& N_1\le \mu\le N,
 \end{array} \right.
\end{equation}
together with
\begin{eqnarray}
  H(\bar{\mathbf s}) &=& H_1( \bar{\mathbf s}^{(1)})+H_2( {\mathbf s}^{(2)}) \\
   &=& H_1( \tilde{\mathbf s}^{(1)})+H_2( {\mathbf s}^{(2)}) \\
   &<&  H_1({\mathbf s}^{(1)})+H_2( {\mathbf s}^{(2)})=H({\mathbf s})
\end{eqnarray}
would yield a state $\bar{\mathbf s}$ with a lower energy than ${\mathbf s}$
which contradicts the assumption that ${\mathbf s}$ is a ground state . Hence ${\mathbf s}^{(1)}$ is a ground state of $H_1$.
The proof that ${\mathbf s}^{(2)}$ is a ground state of $H_2$ is analogous.

It remains to show that ${\mathbf s}$ can be written as a fusion of ${\mathbf s}^{(1)}$  and ${\mathbf s}^{(2)}$.
First we have to define the $N\times 2M$-matrices ${\mathbf S}^{(1)}$ and ${\mathbf S}^{(2)}$ according to (\ref{DF4a}) and (\ref{DF4b}).
The claim of (ii) now follows if there exists a rotation/reflection $R\in O(2M)$ such that $R\,{\mathbf S}^{(2)}_\mu={\mathbf s}^{(2)}_\mu$ for all
$\mu=N_1,\ldots,N$. In the latter equation we have implicitly identified the rows of ${\mathbf s}^{(2)}$
that are vectors ${\mathbf s}^{(2)}_\mu\in {\mathbbm R}^M$ with
the corresponding vectors of ${\mathbbm R}^{2M}$ obtained by padding with $M$ zeroes, compare
the remarks at the beginning of the section about the more precise definition of a state by equivalence classes.
Let ${\mathbf e}_i,\; i=1,\ldots,2M$ denote the standard basis
of ${\mathbbm R}^{2M}$, then the desired $R$ is uniquely defined by
$R\,{\mathbf e}_i ={\mathbf e}_{i+M}$ for $i=1,\ldots,M$ and $R\,{\mathbf e}_i ={\mathbf e}_{i-M}$ for $i=M+1,\ldots,2M$.  \\
 \hfill$\Box$\\

\noindent{\bf Proof of Proposition \ref{PropF2}}\\
We consider the $N\times (M_1+M_2)$-matrices ${\mathbf S}^{(i)},\;i=1,2,$ defined in (\ref{DF4a}) and (\ref{DF4b})
and the orthogonal subspaces $L_i\subset {\mathbbm R}^{M_1+M_2}$ spanned by the rows of ${\mathbf S}^{(i)}$.
Hence $\mbox{dim }L_i=M_i$ for $i=1,2$.
Let $T^{(2)}$ be the orthogonal complement
of the vector ${\mathbf S}^{(2)}_{N_1}$ in $L_2$ such that $\mbox{dim }T^{(2)}=M_2-1$. The rotation $R\in O(M_1+M_2)$ that
maps ${\mathbf S}_{N_1}^{(2)}$ onto ${\mathbf S}_{N_1}^{(1)}$ can be chosen such that it leaves every vector in $T^{(2)}$ fixed.
We have the unique linear decomposition ${\mathbf S}_{\nu}^{(2)}=\alpha_\nu\,{\mathbf S}^{(2)}_{N_1}+\beta_\nu \,{\mathbf t}_\nu$ where
$\nu=N_1+1,\ldots,N$ and ${\mathbf t}_\nu\in T^{(2)}$. This implies
\begin{eqnarray}\label{PF1a}
 R\,{\mathbf S}_{\nu}^{(2)} &=& \alpha_\nu\,R\,{\mathbf S}^{(2)}_{N_1}+\beta_\nu \,R\,{\mathbf t}_\nu \\
 \label{PF1b}
   &=&  \alpha_\nu \,{\mathbf S}^{(1)}_{N_1}+\beta_\nu \,{\mathbf t}_\nu \\
   \nonumber
  \mbox{hence}&&\\
  \nonumber
  \bar{\mathbf S}_\nu^{(2)} &=& R\,{\mathbf S}_\nu^{(2)}\in {\mathbbm R}{\mathbf S}^{(1)}_{N_1} \oplus T^{(2)}\mbox{ for }\nu=N_1,\ldots,N
  \;.\\
  \label{PF1c}
  &&
\end{eqnarray}
Recall that the fusion ${\mathbf s}$ has been defined by
\begin{equation}\label{PF1d}
 {\mathbf s}_\mu \equiv
 \left\{\begin{array}{r@{\quad:\quad}l}
{\mathbf S}_\mu^{(1)}& 1\le \mu,\nu\le N_1,\\
\bar{\mathbf S}_\mu^{(2)}&  N_1\le \mu,\nu\le N.
 \end{array} \right.
\end{equation}
It follows that ${\mathbf s}_\nu \in L_1\oplus T^{(2)}\mbox{ for }\nu=1,\ldots,N$ and hence that $\mbox{dim }{\mathbf s}=M_1+M_2-1$
since the ${\mathbf s}_\nu,\,\nu=1,\ldots,N$ span $L_1\oplus T^{(2)}$.
The fusion ${\mathbf s}$ is a ground state of $H$ by Proposition \ref{PropF1} (i). It remains to show that its dimension is maximal.
In order to derive a contradiction assume that there exists another ground state ${\mathbf S}$ realizing the
maximal dimension $M=\mbox{dim }{\mathbf S}$ of ground states of $H$ and $M>\mbox{dim }{\mathbf s}$. Recall that $M$ can be obtained as the dimension
of the subspace of ${\mathbbm R}^{M_1+M_2}$ spanned by the rows ${\mathbf S}_\mu$ of ${\mathbf S}$. Then there exists a
selection of $M$ linearly independent rows ${\mathbf S}_\mu$ that contains a given row, say, ${\mathbf S}_{N_1}$. In other words,
there exists a subset $F\subset\{1,\ldots,N\}$ of spin numbers such that $N_1\in F$, $|F|=M$ and the set of ${\mathbf S}_\mu,\;\mu\in F,$
is linearly independent. Define $F_1\equiv F\cap \{1,\ldots,N_1\}$, $F_2\equiv F\cap \{N_1,\ldots,N\}$ and
$M^{(i)}\equiv \left|F_i\right|$ for $i=1,2$. Obviously, $F_1\cap F_2=\{N_1\}$ and hence $M=M^{(1)}+M^{(2)}-1$.
As in the proof of Proposition \ref{PropF1} it follows that the states ${\mathbf s}_\mu,\; \mu=1,\ldots,N_1$ and
${\mathbf s}_\mu,\; \mu=N_1,\ldots,N$ are ground states of $H_1$ and $H_2$ with dimension $M^{(i)}$, resp.~.
Since by assumption the $M_i$ are maximal dimensions it follows that $M^{(i)}\le M_i$ for $i=1,2$ and hence
$M\le M_1+M_2-1=\mbox{dim }{\mathbf s}$. The latter contradicts $M>\mbox{dim }{\mathbf s}$ and hence the fusion has maximal dimension
$M=M_1+M_2-1$. This also proves (i).\\

\noindent(ii) $\quad$   We again use a ground state ${\mathbf s}$ realizing the maximal dimension $M$ that can be obtained as the fusion of  ground states
${\mathbf s}^{(1)}$  and ${\mathbf s}^{(2)}$ realizing the maximal dimensions $M_1$ and $M_2$, resp.~. Let ${\mathbf S}^{(1)},\;\bar{\mathbf S}^{(2)}$
be the matrices defined in (\ref{DF4a}) and (\ref{DF6}). Recall that the subspace $P$ of ${\mathcal S}{\mathcal M}(M)$ generated by the projectors $P_\mu,\;\mu=1,\ldots,N$
onto the rows ${\mathbf s}_\mu$ of ${\mathbf s}$ has the dimension $p$. Similarly as above, there exists a selection of spin numbers
$G\subset\{1,\ldots,N\}$ such that $N_1\in G,\;|G|=p$ and the set of projectors $P_\mu,\;\mu\in G$ is linearly independent in ${\mathcal S}
{\mathcal M}(M)$.
Define $G_1\equiv G\cap\{1,\ldots,N_1\},\;G_2\equiv G\cap\{N_1,\ldots,N\}$ and $p^{(i)}=\left| G_i \right|$ for $i=1,2$,
such that $p=p^{(1)}+p^{(2)}-1$. It follows that $p^{(1)}$ and $p^{(2)}$ are also the co-degrees of the matrices
${\mathbf S}^{(1)},$ and $\bar{\mathbf S}^{(2)}$, resp.~. The operations leading from ${\mathbf s}^{(1)}$ to ${\mathbf S}^{(1)}$
and from ${\mathbf s}^{(2)}$ to $\bar{\mathbf S}^{(2)}$, namely padding with zero columns and rotations/reflections, do not change
the co-degree of matrices. Hence $p^{(i)}=p_i$ for $i=1,2,$ and the claim (ii) is proven.\\

\noindent(iii) $\quad$  According to Proposition \ref{PropADE} we have
\begin{equation}\label{PF2}
 d_i= \frac{1}{2}M_i(M_i+1)-p_i,\mbox{ for }i=1,2,
\end{equation}
and hence
\begin{eqnarray}\label{PF3a}
  d &=&\frac{1}{2}\,M(M+1)-p\\
  \nonumber
   &\stackrel{(i)(ii)}{=}& \frac{1}{2}\,(M_1+M_2-1)(M_1+M_2) \\
  \label{PF3b}
   &&-(p_1+p_2-1)\\
   \nonumber
  &=& \frac{1}{2}\,(M_1^2+M_2^2-M_1-M_2+2M_1M_2)\\
  \label{PF3c}
  &&-p_1-p_2+1\\
  \nonumber
  &=& \left(\frac{1}{2}M_1(M_1+1)-p_1\right)\\
  \nonumber
  &&+\left(\frac{1}{2}M_2(M_2+1)-p_2\right)\\
   \label{PF3d}
  &&-M_1-M_2+M_1 M_2+1 \\
  &\stackrel{(\ref{PF2})}{=}&d_1+d_2+(M_1-1)(M_2-1)
  \;.
\end{eqnarray}
This completes the proof of Proposition \ref{PropF2}. \hfill$\Box$\\

\subsection{Proof of Theorem \ref{Theorem1}}\label{sec:PT1}

It turns out that the proof of  Theorem \ref{Theorem1} considerably simplifies for
regular points of ${\mathcal V}$.  Hence we will treat this special case in a separate subsection.

\subsubsection{The regular case}\label{sec:PT1R}

If $({\boldsymbol\lambda},x)$ is a regular point of ${\mathcal V}$ we have already shown that the upper
cone degenerates into a half-space,  $ {\mathcal C}^+({\boldsymbol\lambda},x)= H^+_\varphi,\;\varphi\in S_1$.
Hence $({\boldsymbol\lambda},x)$ is vertical iff $H^+_\varphi=H^+$.

For the only-if-part of the theorem assume
that there is a state ${\mathbf s}$ living on the one-dimensional eigenspace $S$ of $({\mathbbm J}({\boldsymbol\lambda}),x)$.
Then ${\mathbf s}$ is necessarily collinear and ${\mathbf s}_\lambda^2={\mathbf s}_N^2=1$ for all $\lambda=1,\ldots,N-1$.
We may set ${\mathbf s}=\sqrt{N}\varphi$. Fix any $\lambda=1,\ldots,N-1$ and consider ${\boldsymbol\mu}\in\Lambda$  of the form
\begin{equation}\label{PT1R1}
 {\mu}_\nu=\delta_{\lambda\nu}-\delta_{N\nu},\;\nu=1,\ldots,N
 \;.
\end{equation}
 Then $\langle \varphi|{\mathbf D}\cdot{\boldsymbol\mu}|\varphi\rangle=\varphi_\lambda^2-\varphi_N^2=0$.
Since all ${\boldsymbol\mu}\in\Lambda$ of the kind (\ref{PT1R1}), $\lambda=1,\ldots,N-1$ form a basis of $\Lambda$ the equation
$\langle \varphi|{\mathbf D}\cdot{\boldsymbol\mu}|\varphi\rangle=0$ holds for all ${\boldsymbol\mu}\in\Lambda$.
Using (\ref{SP3}) it follows that the graph of $h_\varphi$ consists of the hyperplane
$H^0\equiv \{ ({\boldsymbol\mu},0)|{\boldsymbol\mu}\in\Lambda\}$ and hence the super-graph $H^+_\varphi$ of $h_\varphi$
will be $H^+$.

For the if-part of the theorem assume that $({\boldsymbol\lambda},x)$ is a vertical point of ${\mathcal V}$,
i.~e.~, $H^+_\varphi=H^+$ for $\varphi\in S_1$. As in the previous paragraph it follows that the graph of $h_\varphi$ consists of the hyperplane
$H^0$ and hence $\varphi_\lambda^2-\varphi_N^2=0$ for all $\lambda=1,\ldots,N-1$. Hence $\sqrt{N}\varphi$ is a collinear
state living on $S$.

This completes the proof of Theorem \ref{Theorem1} for regular points of ${\mathcal V}$.

\subsubsection{The singular case, only-if-part of Theorem \ref{Theorem1}}\label{sec:PT1Sif}

For the only-if-part of Theorem \ref{Theorem1} we will assume that $({\boldsymbol\lambda},x)$ is an elliptic point of ${\mathcal V}$.
According to Proposition \ref{PropC} the eigenspace $S'$ of $({\mathbbm J}({\boldsymbol\lambda}),x)$ will contain a completely elliptic subspace $S$.
The restriction to $\varphi\in S_1$ possibly enlarges ${\mathcal C}^+({\boldsymbol\lambda},x)$ to some cone
$\widetilde{\mathcal C}^+({\boldsymbol\lambda},x)$. If $\widetilde{\mathcal C}^+({\boldsymbol\lambda},x)$ is shown to be vertical, also
the sub-cone ${\mathcal C}^+({\boldsymbol\lambda},x)$ will be vertical and hence it will be sufficient to work with the completely elliptic subspace $S$.

It then follows that the ADE (\ref{DSG6}) has a solution $\Delta>0$  yielding spin configurations
\begin{equation}\label{if1}
{\mathbf s}=W\,\sqrt{\Delta}\,R
\;,
\end{equation}
where the columns of the $N\times M$-matrix $W$ will be assumed to span an orthonormal basis of the $M$-dimensional subspace $S$
and $R\in O(M)$ is arbitrary. If $M=1$ we will proceed as in the only-if-part of the regular case in subsection \ref{sec:PT1R}.
Hence we may assume that $M>1$ in what follows.

Now consider ${\mathbf s}^\top\,{\mathbf s}= R^\top\,\sqrt{\Delta}\,W^\top\,W\,\sqrt{\Delta}\,R=R^\top\,\Delta\,R$.
Here we have used $W^\top\,W={\mathbbm 1}_M$ since the columns of $W$ are orthonormal. We choose $R\in O(M)$ such that
$R^\top\,\Delta\,R$ becomes a diagonal matrix, say, ${\mathbf s}^\top\,{\mathbf s}=R^\top\,\Delta\,R=\mbox{diag }(\delta_1,\ldots,\delta_M)$.
The latter equation says that $\left(\delta_i^{-1/2}\, {\mathbf s}_i\right)_{i=1,\ldots,M}$ will
be an orthonormal basis of $S$.

Let ${\mathbbm Q}$ denote the projector onto $S$. For any given ${\boldsymbol\mu}\in\Lambda$ let
$a_{min}$ and $a_{max}$ denote the lowest and highest eigenvalue of ${\mathbbm Q}\,{\mathbf D}\cdot {\boldsymbol\mu}\,{\mathbbm Q}$.
Any convex combination of the expectation values
$\alpha_i\equiv\langle \delta_i^{-1/2}\, {\mathbf s}_i| {\mathbf D}\cdot {\boldsymbol\mu}|\delta_i^{-1/2}\, {\mathbf s}_i\rangle$
lies in the interval $[a_{min},a_{max}]$, hence
\begin{equation}\label{if2}
 a_{min}\le \,\frac{1}{\mbox{Tr }\Delta}\sum_{i=1}^M\delta_i\,\alpha_i\,\le a_{max}
 \;.
\end{equation}
The sum in (\ref{if2}) is evaluated as follows:
\begin{eqnarray}\label{if3}
  \sum_{i=1}^M\delta_i\,\alpha_i &=&\sum_{i=1}^M \langle {\mathbf s}_i| {\mathbf D}\cdot {\boldsymbol\mu}|{\mathbf s}_i\rangle \\
  \label{if4}
   &=& \sum_{\lambda,i} {\mathbf s}_{\lambda i}^2\, {\mu}_\lambda =
   \sum_{\lambda=1}^N \left( \sum_{i=1}^M   {\mathbf s}_{\lambda i}^2\right)\,{\mu}_\lambda \\
   \label{if5}
    &=& \sum_{\lambda=1}^N {\mu}_\lambda=0
    \;.
\end{eqnarray}
(\ref{if2}) and (\ref{if5}) imply
\begin{equation}\label{if6}
 a_{min}\le 0 \le a_{max}
 \;,
\end{equation}
and hence $({\boldsymbol\lambda},x)$ will be a vertical point of ${\mathcal V}$, see (\ref{SP14}).
This completes the proof of the only-if-part of Theorem {\ref{Theorem1} in the singular case.

\subsubsection{The singular case, if-part of Theorem \ref{Theorem1}}\label{sec:PT1Sonlyif}

We will use some elementary notions of convex analysis that can be found, e.~g., in \cite{R97}.
Recall that $\Lambda\times {\mathbbm R} \cong {\mathbbm R}^N$. Instead of $({\boldsymbol\lambda},x)$ it is sometimes
more convenient to use the new coordinates ${\boldsymbol\kappa}\in {\mathbbm R}^N$ for $\Lambda\times {\mathbbm R}$ that are related to the old ones by
\begin{equation}\label{oi3}
{\boldsymbol\kappa}={\boldsymbol\lambda}+\bar{\kappa}\,{\mathbf e}={\boldsymbol\lambda}-x\,{\mathbf e}
\;,
\end{equation}
where
\begin{equation}\label{oi4}
{\mathbf e}\equiv(1,1,\ldots,1)^\top\in {\mathbbm R}^N
\;.
\end{equation}
This entails some minor modifications of the definitions concerning the Lagrange variety etc.~,
but these modifications will only be valid for this subsection. First, we re-define the dressed ${\mathbbm J}$-matrix
and the Lagrange variety according to
\begin{equation}\label{oi1}
 {\mathbbm J}({\boldsymbol\kappa})\equiv{\mathbbm J}+{\mathbf D}\cdot{\boldsymbol\kappa},\;{\boldsymbol\kappa}\in{\mathbbm R}^N
 \;,
\end{equation}
and
\begin{equation}\label{oi2}
 {\mathcal V}\equiv \{ {\boldsymbol\kappa}\in {\mathbbm R}^N\left| \det {\mathbbm J}({\boldsymbol\kappa})=0\right.\}
 \;.
\end{equation}

For any  ${\boldsymbol\kappa}\in{\mathcal V }$ let $S$ denote the null space of ${\mathbbm J}({\boldsymbol\kappa})$ and
$S_1$, as before, the subset of unit vectors. Further, let ${\mathcal W}_+(S)$ denote the closed convex cone
of all real symmetric positively semi-definite operators $W:S\longrightarrow S$. Further we consider the closed convex cone
\begin{equation}\label{oi5}
  B\equiv \{ \mbox{Tr }(W {\mathbf D}) | W\in {\mathcal W}_+(S)\}\subset {\mathbbm R}^N
  \;.
\end{equation}

For $\varphi\in S_1$ re-define the closed upper half-space by
$H_\varphi^+ \equiv \{ {\boldsymbol\alpha}\in {\mathbbm R}^N | \langle \varphi|{\mathbf D}\cdot{\boldsymbol\alpha}|\varphi\rangle\le 0\}$.
Note that the $\le$ is not a typo but results from (\ref{oi3}) and the requirement of consistency with (\ref{SP3}).
In accordance to the previous definitions we set ${\mathcal C}^+({\boldsymbol\kappa})\equiv\bigcap_{\varphi\in S_1}H_\varphi^+ \subset {\mathbbm R}^N$.
${\mathcal C}^+({\boldsymbol\kappa})$ is called ``vertical" iff ${\boldsymbol\alpha}\cdot {\mathbf e}\le 0$ for all
${\boldsymbol\alpha}\in {\mathcal C}^+({\boldsymbol\kappa})$.

For any closed convex cone $K\subset {\mathbbm R}^N$ we will consider the closed convex ``dual cone"
$K^\ast\equiv \{ {\boldsymbol\beta}\in {\mathbbm R}^N |  {\boldsymbol\beta} \cdot  {\boldsymbol\alpha}\ge 0 \mbox{ for all } {\boldsymbol\alpha}\in K\}$.
Inclusion of cones is reversed by duality: $K_1\subset K_2$ implies $K_2^\ast\subset K_1^\ast$.
According to a general theorem, $K^{\ast\ast}=K$, see \cite{R97}, Theorem 14.5.
The condition of  ${\mathcal C}^+({\boldsymbol\kappa})$ being vertical now can be reformulated as ${\mathcal C}^+({\boldsymbol\kappa})\subset -E^\ast$
where $E$ denotes the closed convex cone $E\equiv \{\alpha\, {\mathbf e}\,|\,\alpha\ge 0\}$.

Consider the following equivalences
\begin{eqnarray}\label{oi6}
   {\boldsymbol\alpha}\in C\equiv{\mathcal C}^+({\boldsymbol\kappa})
   &\Leftrightarrow&
    {\boldsymbol\alpha}\in H_\varphi^+  \quad\forall\,\varphi\in S_1\\
    \label{oi7}
         &\Leftrightarrow&
    \langle \varphi | {\mathbf D}\cdot{\boldsymbol\alpha}|\varphi\rangle\le 0\quad \forall\, \varphi\in S_1\\
    \nonumber
     &\Leftrightarrow&
     \mbox{Tr }(W  {\mathbf D}\cdot{\boldsymbol\alpha})\le 0 \quad\forall\, W\in {\mathcal W}_+(S)\\
     \label{oi8}
     &&\\
     \nonumber
      &\Leftrightarrow&
       \mbox{Tr }(W  {\mathbf D})\cdot{\boldsymbol\alpha}\le 0 \quad\forall\, W\in {\mathcal W}_+(S)\\
       \label{oi9}
       &&\\
        \label{oi10}
       &\Leftrightarrow&
       {\boldsymbol\alpha}\in - B^\ast
       \;,
   \end{eqnarray}
where the equivalence (\ref{oi8}) follows by the spectral theorem. We thus proved $C =-B^{\ast}$ and hence
\begin{equation}\label{oi11}
 C^\ast = - B^{\ast\ast} = -B
 \;.
\end{equation}
Now we conclude
\begin{eqnarray}\label{oi12}
  C\mbox{ vertical} &\Leftrightarrow& C\subset -E^\ast \\
  \label{oi13}
  &\Leftrightarrow& E\subset - C^\ast \\
  \label{oi14}
  &\stackrel{(\ref{oi11})}{\Leftrightarrow}& E\subset B \\
  &\stackrel{(\ref{oi5})}{\Leftrightarrow}& \exists\, W\in {\mathcal W}_+(S): \mbox{ Tr} (W{\mathbf D})={\mathbf e}\\
  \nonumber
   &\Leftrightarrow& \exists\, W\in {\mathcal W}_+(S): \mbox{ Tr} (W{D}_\mu)=1\\
   \label{oi15}
   &&\forall \mu=1,\ldots,N.
    \end{eqnarray}
Let the rank of $W$ be $m$ and $W=\sum_{i=1}^m w_i \,{\mathbbm P}_{\psi_i}$ be the spectral decomposition of $W\in {\mathcal W}_+(S)$ such that
$w_i>0$ for $i=1,\ldots,m$. Then for all $\mu=1,\ldots,N$ we can evaluate (\ref{oi15}) as follows
\begin{eqnarray}\label{oi16}
  1&=&\mbox{ Tr} (W{D}_\mu) \\
  \label{oi17}
   &=& \sum_{i=1}^m w_i \langle \psi_i|D_\mu|\psi_i\rangle \\
   \label{oi18}
   &=& \sum_{i=1}^m w_i \psi_{i\mu}^2 = \sum_{i=1}^m {\mathbf s}_{\mu i}^2
      \;,
\end{eqnarray}
where we have set ${\mathbf s}_{\mu i}\equiv \sqrt{w_i}\,\psi_{i\mu}$. This proves that for any vertical
${\mathcal C}^+({\boldsymbol\kappa})$ there exists an $m$-dimensional spin configuration ${\mathbf s}$
living on the corresponding eigenspace $S$.

\subsection{Existence and uniqueness of ground states}\label{sec:PEU}
\noindent{\bf Proof of Theorem \ref{Theorem3}}\\
In order to derive a contradiction, let us assume that there exist
${\boldsymbol\lambda}^{(1)},\,{\boldsymbol\lambda}^{(2)}\in\widehat{J}$ such that ${\boldsymbol\lambda}^{(1)}\neq{\boldsymbol\lambda}^{(2)}$.
By convexity of $\widehat{J}$ it follows that also
\begin{equation}\label{PEU1}
 {\boldsymbol\lambda}\equiv \frac{1}{2}\left({\boldsymbol\lambda}^{(1)}+{\boldsymbol\lambda}^{(2)} \right)\in\widehat{J}
 \;.
\end{equation}
Then, for  $|\epsilon|\le 1$,
\begin{equation}\label{PEU2}
j_{min }\left({\boldsymbol\lambda}+\epsilon \left({\boldsymbol\lambda}^{(1)}-{\boldsymbol\lambda} \right)\right)=\hat{\jmath}
\;.
\end{equation}
Let $S$ be the eigenspace of $({\mathbbm J}({\boldsymbol\lambda}),\hat{\jmath})$, and ${\mathbbm Q}$ the projector onto $S$.
According to degenerate perturbation theory the eigenvalue $j_{min}({\boldsymbol\lambda})$ will split into $n$ possibly different
eigenvalues $x_i(\epsilon)$ such that $x_i(0)=j_{min}({\boldsymbol\lambda})=\hat{\jmath}$ and
\begin{equation}\label{PEU4}
x_i(\epsilon)=\hat{\jmath} + \epsilon \left\langle \varphi_i\left| {\mathbf D}\cdot \left({\boldsymbol\lambda}^{(1)}-{\boldsymbol\lambda} \right)
\right| \varphi_i\right\rangle + {\mathcal O}(\epsilon^2)
\;,
\end{equation}
where $|\epsilon|$ is sufficiently small and the $\varphi_i,\,i=1,\ldots,n$ are the eigenvectors of
${\mathbbm Q}\,{\mathbf D}\cdot \left({\boldsymbol\lambda}^{(1)}-{\boldsymbol\lambda} \right)\,{\mathbbm Q}$.
The two equations (\ref{PEU2}) and (\ref{PEU4}) are only compatible if
${\mathbbm Q}\,{\mathbf D}\cdot \left({\boldsymbol\lambda}^{(1)}-{\boldsymbol\lambda} \right)\,{\mathbbm Q}=0$, i.~e.~, if
\begin{equation}\label{PEU5}
   \left\langle \varphi\left| {\mathbf D}\cdot \left({\boldsymbol\lambda}^{(1)}-{\boldsymbol\lambda} \right)
\right| \varphi\right\rangle=0
\end{equation}
for all $\varphi\in S_1$. It follows that
\begin{eqnarray}\nonumber
\left\langle \varphi\left| {\mathbbm J}\left({\boldsymbol\lambda}^{(1)}\right)\right| \varphi\right\rangle
 &=& \left\langle \varphi\left|
 {\mathbbm J}\left({\boldsymbol\lambda}+\left({\boldsymbol\lambda}^{(1)}-{\boldsymbol\lambda}\right)\right)
 \right| \varphi\right\rangle\\
\nonumber
 &&\\
 \nonumber
 &=&\left\langle \varphi\left|  {\mathbbm J}\left({\boldsymbol\lambda}\right) \right|\varphi \right\rangle\\
  &+&
 \left\langle \varphi\left|  {\mathbf D}\cdot\left({\boldsymbol\lambda}^{(1)}-{\boldsymbol\lambda}\right) \right|\varphi \right\rangle \\
 &\stackrel{(\ref{PEU5})}{=}&
 \left\langle \varphi\left|  {\mathbbm J}\left({\boldsymbol\lambda}\right) \right|\varphi \right\rangle=\hat{\jmath}
 \;.
\end{eqnarray}
Hence all $\varphi\in S_1$ realize the minimal expectation value
$\hat{\jmath}$ of ${\mathbbm J}\left({\boldsymbol\lambda}^{(1)}\right)$
and consequently must be eigenvectors of $\left({\mathbbm J}\left({\boldsymbol\lambda}^{(1)}\right),\hat{\jmath}\right)$.
Let $S^{(1)}$ denote the
eigenspace of $\left({\mathbbm J}\left({\boldsymbol\lambda}^{(1)}\right),\hat{\jmath}\right)$. We thus have shown $S\subset S^{(1)}$.
It follows that any ground state ${\mathbf s}$ that lives on $S$ also lives on $S^{(1)}$. Since the Lagrange parameters
${\boldsymbol\kappa}$ only depend on ${\mathbf s}$ it follows further that ${\boldsymbol\lambda}={\boldsymbol\lambda}^{(1)}$
which contradicts ${\boldsymbol\lambda}^{(1)}\neq{\boldsymbol\lambda}^{(2)}$.                            \hfill$\Box$\\

\noindent{\bf Proof of Theorem \ref{TheoremSym}}\\
The proof anticipates some notions of the Gram set approach. The ``Gram set" is the convex set of Gram matrices defined by
\begin{equation}\label{sym2}
{\mathcal G}\equiv \left\{G\in {\mathcal S}{\mathcal M}_+(N)\left| G_{\mu \mu}=1 \mbox{ for all }\mu=1,\ldots, N\right.\right\}
\;.
\end{equation}
It contains the non-empty convex subset (actually a face of ${\mathcal G}$)
\begin{equation}\label{sym2}
\breve{\mathcal G}\equiv \left\{G\in {\mathcal G}\left|\mbox{ Tr }\left( G\,{\mathbbm J}\right)=E_{\min}\right.\right\}
\end{equation}
of Gram matrices corresponding to ground states. Let $G\in \breve{\mathcal G}$, then the Gram matrix
\begin{equation}\label{sym3}
\dot{G}\equiv \frac{1}{|{\sf Gr}|}\sum_{\Pi\in{\sf Gr}}\,\Pi\,G\,\Pi^\top
\end{equation}
satisfies $\dot{G}\in \breve{\mathcal G} $ and is obviously invariant under the action of ${ \sf Gr}$, hence it is the Gram matrix of a symmetric ground state.
 \hfill$\Box$\\

We note that symmetric ground states can be calculated by group-theoretical means, similarly as already considered in \cite{SL03}. The subspace
$S$, being invariant under all $\Pi\in{\sf Gr}$, can be decomposed into irreducible representations of ${\sf Gr}$ with projections
${\mathbbm Q}_i,\,i=1,\ldots k$ that are, however, not unique in general. 
Since the Gram matrix $G$ of a symmetric ground state commutes with all $\Pi\in{\sf Gr}$ it must be
a non-negative linear combination of certain suitable ${\mathbbm Q}_i,\,i=1,\ldots k$, by virtue of Schur's lemma. If we restrict ${\sf Gr}$ to
an Abelian subgroup ${\mathcal T}$ of translations it follows that all real irreducible representations of ${\mathcal T}$ are at most
two-dimensional. Hence Theorem \ref{TheoremSym} implies the existence of ${\mathcal T}$-symmetric ground states that are co-planar or
collinear.

Further investigations of symmetric ground states have to be deferred to subsequent papers.

\section{Summary}\label{sec:SUM}
We will summarize the central results of this paper in a theorem that contains also the pertinent definitions and can be read
independently of the main text.
\begin{theorem}\label{TSUM}
For all integers $N\ge 2$ and $M\ge 1$
let ${\mathcal P}_M$ denote the set of real $N\times M$-matrices ${\mathbf s}$ such that the $N$ rows ${\mathbf s}_\mu$ of ${\mathbf s}$
satisfy
\begin{equation}\label{TSUM1}
  {\mathbf s}_\mu\cdot  {\mathbf s}_\mu=1 \mbox{ for } \mu=1,\ldots,N
  \;,
\end{equation}
and let the Heisenberg spin system be characterized by its Hamiltonian
\begin{equation}\label{TSUM2}
 H({\mathbf s})=\sum_{\mu,\nu=1}^{N}J_{\mu\nu}\, {\mathbf s}_\mu\cdot  {\mathbf s}_\nu
  \;.
\end{equation}
Let $E_{min}=\mbox{Min }\left\{ H({\mathbf s})\left|{\mathbf s}\in{\mathcal P}_N \right. \right\} $
and $\breve{\mathcal P}=\left\{{\mathbf s}\in{\mathcal P}_N\left|H({\mathbf s})=E_{min} \right. \right\}$
denote the set of ground states of (\ref{TSUM2}). Without loss of generality we may assume
$\breve{\mathcal P}\subset{\mathcal P}_{\breve{M}}$ where $\breve{M}$ denotes the maximal rank of ground states.\\

For all ${\boldsymbol\lambda}\in{\mathbbm R}^N$ satisfying
\begin{equation}\label{TSUM3}
\sum_{\mu=1}^{N}\lambda_\mu=0  \;,
\end{equation}
we define ${\mathbbm J}({\boldsymbol\lambda})$ as the real, symmetric $N\times N$-matrix with entries
\begin{equation}\label{TSUM4}
 J_{\mu\nu}({\boldsymbol\lambda})=\left\{\begin{array}{l@{\quad:\quad}l}
                                           J_{\mu\nu} & \mu\neq \nu  \\
                                           \lambda_\mu & \mu = \nu,
                                         \end{array}
  \right.
\end{equation}
and denote by $j_{min}({\boldsymbol\lambda})$ its lowest eigenvalue.\\
Then there exists a unique $\hat{\boldsymbol\lambda}$ where $j_{min}({\boldsymbol\lambda})$ assumes its
maximum $\hat{\jmath}=j_{min}(\hat{\boldsymbol\lambda})$. Let $S$ be the corresponding eigenspace of
${\mathbbm J}(\hat{\boldsymbol\lambda})$.
Any ground state ${\mathbf s}\in\breve{\mathcal P}$ will be of the form
\begin{equation}\label{TSUM5}
 {\mathbf s}=W\,\sqrt{\Delta}\,R
 \;,
\end{equation}
where $W$ is an $N\times \breve{M}$-matrix the columns of which span $S$, $\Delta$ is a positively semi-definite  $\breve{M}\times \breve{M}$-matrix,
and $R\in O(\breve{M})$ an $\breve{M}$-dimensional rotation or reflection.
Moreover, the set of $\Delta\ge 0$ such that (\ref{TSUM5}) defines a ground state of $H$ is an $d$-dimensional compact convex set
characteristic for the spin system under consideration.
\end{theorem}

\section*{Acknowledgment}
I have greatly profited from the long lasting cooperation with Marshall Luban and Christian Schr\"oder including work
on classical ground states that has left its mark on the theory presented here. Moreover, I gratefully acknowledge
discussions with Thomas Br\"ocker, Johannes Richter and J\"urgen Schnack on the very subject.


\end{document}